\documentclass[superscriptaddress, aps,prb,twocolumn,longbibliography, floatfix]{revtex4-2}

\usepackage{amsmath}
\usepackage{amsfonts}
\usepackage{amsthm}
\usepackage{amssymb}
 
\usepackage{bm}
\usepackage{physics}

\usepackage{graphicx}
\usepackage{xcolor}
\usepackage[caption=false]{subfig}

\usepackage[breaklinks=true,colorlinks,citecolor=blue,linkcolor=red,urlcolor=blue]{hyperref}

\newcommand{\dt}{\frac{d}{dt}}
\newcommand{\ii}{\rm{i}}
\newcommand{\ee}{\rm{e}}
\newcommand{\ce}{\varepsilon}

\newcommand{\fig}{Fig.}

\newcommand{\eq}{Eq.}

\renewcommand{\rm}[1]{\mathrm{#1}}
\renewcommand{\dag}{\dagger}

\begin{document}
\title{Variational approach to the dynamics of dissipative quantum impurity models}

\author{Yi-Fan Qu}
\email{yifaqu@phys.ethz.ch}
\affiliation{Institute for Theoretical Physics, ETH Zurich, 8093 Zurich, Switzerland}

\author{Martino Stefanini}
\email{mstefa@uni-mainz.de}
\thanks{Y.-F. Q. and M. S. contributed equally to this work.}
\affiliation{Institut f\"ur Physik, Johannes Gutenberg-Universit\"at Mainz, D-55099 Mainz, Germany}


\author{Tao Shi}
\affiliation{CAS Key Laboratory of Theoretical Physics, Institute of Theoretical Physics, Chinese Academy of Sciences, Beijing 100190, China}
\affiliation{CAS Center for Excellence in Topological Quantum Computation \& School of Physical Sciences, University of Chinese Academy of Sciences, Beijing 100049, China}

\author{Tilman Esslinger}
\affiliation{Institute for Quantum Electronics \& Quantum Center, ETH Zurich, 8093 Zurich, Switzerland}

\author{Sarang Gopalakrishnan}
\affiliation{Department of Electrical and Computer Engineering, Princeton University Princeton, New Jersey 08540, USA}

\author{Jamir Marino}
\affiliation{Institut f\"ur Physik, Johannes Gutenberg-Universit\"at Mainz, D-55099 Mainz, Germany}

\author{Eugene Demler}
\affiliation{Institute for Theoretical Physics, ETH Zurich, 8093 Zurich, Switzerland}

\date{\today}

\begin{abstract}
Recent experiments with quantum simulators using ultracold atoms and superconducting qubits have demonstrated the potential of controlled dissipation as a versatile tool for realizing correlated many-body states. However, determining the dynamics of dissipative quantum many-body systems remains a significant analytical and numerical challenge. In this work, we focus on a dissipative impurity problem as a testbed for new methodological developments. We introduce an efficient non-perturbative framework that combines the superposition of Gaussian states (SGS) variational ansatz with the quantum trajectory approach to simulate open systems featuring a dissipative impurity.
Applying this method to a spinful impurity subject to two-body losses and embedded in a bath of noninteracting fermions, we explore the full crossover from weak to strong dissipation regimes. The non-perturbative nature of the SGS ansatz allows us to thoroughly examine this crossover, providing comprehensive insights into the system's behavior. In the strong dissipation regime, our approach reproduces the finding that localized two-body losses can induce the Kondo effect [arXiv:2406.03527], characterized by a slowdown of spin relaxation and an enhancement of charge conductance. Furthermore, we reveal an exotic ``negative conductance" phenomenon at zero potential bias -- a counter-intuitive single-body effect resulting from intermediate dissipation and finite bandwidth.
Finally, we investigate the formation of ferromagnetic domains and propose an extension to realize a higher-spin Kondo model using localized dissipation.

\end{abstract}

\maketitle

\section{Introduction}

Dissipative quantum physics has been an active area of research for several decades. Coupling a quantum system to an environment not only acts detrimentally by inducing decoherence and dephasing, but can also result in the onset of phase transitions, as in the celebrated case of the spin-boson model \cite{Leggett1987}. With the advent of ultracold quantum simulators, it has been realized that dissipation can be used as a tool for preparing correlated states of matter \cite{kraus2008,diehl2008quantum,verstraete2009,lin2013,bardyn2013,shankar2013,Nakagawa2020,raghunandan2020,sieberer2023universality} (see also recent work on quantum simulators using superconducting qubits \cite{ma2019,Mi2024}).
These developments underscore that dissipation, while typically suppressing quantum coherence, can also give rise to novel phenomena without counterparts in unitary systems, such as non-equilibrium critical steady states \cite{dalla2010,dallatorre2012} and instances of quantum many-body Zeno physics \cite{misra1977,fisher2001,facchi2002,syassen2008,facchi2008,Sun2023}.

Understanding dissipative systems poses significant analytical and numerical challenges. Various simulation methods have been employed, including perturbative techniques like the Keldysh formalism \cite{Sieberer2013,torre2013,Sieberer2016,Maghrebi2016,gievers2023} and non-perturbative approaches such as exact diagonalization (ED) \cite{Cai2013,PhysRevE.92.042143,Nakagawa2020}, scattering theory \cite{rephaeli2013,zhou2021}, density matrix renormalization group (DMRG) \cite{Daley2009,bonnes2014,yamamoto2022}, matrix product operators (MPOs) \cite{Mascarenhas2015,Cui2015}, Bethe ansatz \cite{buca2020,nakagawa2021,alba2023,ekman2024}, and variational approaches \cite{hartmann2019,Nagy2019,Vicentini2019, bond2024, Mazza2023}.
However, non-perturbative methods face significant challenges when applied to large non-integrable dissipative systems, since the dimension of the space of density matrices grows with the size of the system even faster than for pure states. Thus, an important goal of theoretical quantum physics has been the development of approximations that are still non-perturbative but are less computationally intensive.

A class of dissipative systems garnering increasing attention is that of dissipative impurities \cite{tonielli2019orthogonality,Ott_dissipative_imp,Dolgirev,tonielli2020ramsey,chaudhari2022zeno,gievers2023},
where dissipation is confined to a small region within a larger quantum system. As their unitary counterparts (such as the Kondo \cite{Kondo}
and Anderson impurity models \cite{anderson1961} and polarons \cite{Landau1933,pekar1946}, to mention only a few), they represent an ideal playground to understand the effect of dissipation in a simplified setting.
In this article, we present a non-perturbative framework that combines the superposition of Gaussian states (SGS) variational ansatz---which is efficient for impurity problems \cite{BravyiGosset_complexity,boutin2021quantum,nondissipativeSGS}---with the quantum trajectory approach \cite{Dalibard1992,Molmer1993,Dum1992,Dum1992_2,carmichael1993open}.
This framework significantly accelerates numerical simulations the dynamics in dissipative impurity systems, extending the accessible system size from $l_\rm{sys}\lesssim20$ (typically with DMRG \cite{bonnes2014,Nakagawa2020} and neural network approaches \cite{hartmann2019}) to $l_\rm{sys} \simeq 160$ using the SGS representation, where each spinful fermionic site is counted as two sites. This scalability makes it a promising tool for studying other platforms where local dissipation acts on an extended quantum system.

We apply our method to a dissipative impurity system where a spinful impurity, immersed in a bath of noninteracting fermions, undergoes two-body losses. Our approach effectively demonstrates that strong localized two-body losses give rise to the Kondo effect \cite{Hewson}, in agreement with the analysis presented in the companion paper \cite{short_paper}. We characterize the emergence of Kondo physics by identifying two distinctive signatures: the slowdown of spin relaxation and the enhancement of charge conductance. Owing to the non-perturbative nature of the SGS ansatz, our method allows for a thorough examination of the full crossover from the weakly dissipative, mean-field regime to the strongly correlated Kondo regime, providing a comprehensive understanding of the system's behavior across different dissipation regimes. Notably, in the mean-field regime, we reveal the emergence of exotic ``negative conductance" at zero potential bias, arising from intermediate dissipation and the constraints of a finite reservoir bandwidth.
Furthermore, we explore a straightforward extension of the dissipative setup by introducing dissipation to multiple sites, enabling the realization of a spin-$1$ Kondo model. This development opens up new avenues for exploring higher-spin Kondo physics utilizing dissipation.

The paper is organized as follows. In Sec.~\ref{sec: model}, we present the dissipative impurity model with two-body losses. 
Sec.~\ref{sec: method} introduces the SGS framework for simulating open systems, which is combined with the quantum trajectory approach.
In Sec.~\ref{sec: results}, we first benchmark our approach using ED. We then investigate the crossover to Kondo regimes through spin relaxation and conductance. In Sec.~\ref{sec: negative conductance}, we demonstrate the origin of the ``negative conductance" phenomenon at zero potential bias.
In Sec.~\ref{sec: high spin} we discuss the formation of ferromagnetic domains and the realization of high-spin Kondo models.

\section{Impurity model} \label{sec: model}
\begin{figure}
    \centering
    \includegraphics[width=0.9\linewidth]{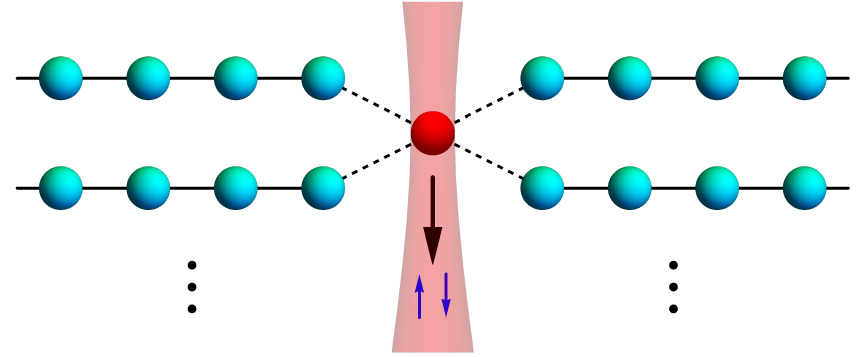}
    \caption{Schematic of the setup with two-body dissipation at an impurity site, which is immersed in a spinful fermionic bath consisting of multiple channels. }
    \label{fig: setup}
\end{figure}

We consider an impurity site immersed in a bath of noninteracting spinful fermions. As illustrated in Fig.~\ref{fig: setup}, fermions on the central impurity site, can undergo a two-body loss: when two fermions with opposite spins occupy this site, they can be lost from the system at a rate $\gamma$. Meanwhile, they can also tunnel into the bath, which consists of multiple channels (depicted by the cyan balls in Fig.~\ref{fig: setup}). Assuming the losses are Markovian \cite{DipolarMolecules2BLoss, PRL_2BLoss_Zeno, DissipativeFermiHubbard,Esslinger-PhysRevA.100.053605}, the dynamics of the density matrix $\hat{\rho}(t)$ of the system is described by a Lindblad master equation
\begin{equation} \label{eq: master equation}
    \frac{d}{dt}\hat{\rho}(t)=-i \comm{\hat{H}}{\hat{\rho}(t)}+\gamma\Big(\hat{L}\hat{\rho}(t)\hat{L}^\dag-\frac{1}{2}\{\hat{L}^\dag \hat{L},\hat{\rho}(t)\}\Big),
\end{equation}
where the unitary part of the dynamics is governed by the Hamiltonian
\begin{equation}\label{eq: model}
    \begin{aligned}
        \hat{H}=&\hat{H}_d+\hat{H}_\textup{leads}+\hat{H}_\textup{tun} \\
        =&\ce_d\sum_\sigma \hat{d}_{\sigma}^\dag \hat{d}_{\sigma}+\sum_{p\sigma\alpha} \ce_{p\alpha} \hat{c}_{p\sigma\alpha}^\dag \hat{c}_{p\sigma\alpha} \\
    &+\sum_{p\sigma\alpha}(V_{p\alpha} \hat{d}_{\sigma}^\dag \hat{c}_{p\sigma\alpha}+\rm{H.c.}).
    \end{aligned}
\end{equation}
Here, the $\hat{d}_{\sigma}$ operators represent the modes of spin $\sigma$  with onsite energy $\ce_d$ at the impurity site. In the spinful fermionic bath, the $\hat{c}_{p\sigma\alpha}$ operators correspond to the eigenmodes of channel $\alpha$, which have single-particle energies $\ce_{p\alpha}=\ce_p-\mu_\alpha$ with momentum $p$ and chemical potential $\mu_\alpha$. Fermions at the impurity site can tunnel into the bath via the hopping term $\hat{H}_\rm{tun}$ with channel-dependent amplitudes $V_{p \alpha}$.
The dissipative part of the dynamics in Eq.~\eqref{eq: master equation} is described by the jump operator $\hat{L}= \hat{d}_{\downarrow} \hat{d}_{\uparrow}$, which represents two-body loss on the impurity site. 

It is worth noting that our model is related to the well-known Anderson impurity model (AIM) described by the Hamiltonian
\begin{equation} \label{eq: AIM}
    \hat{H}_\rm{AIM}=\ce_d\sum_\sigma \hat{d}_{\sigma}^\dag \hat{d}_{\sigma} + \frac{U_d}{2} \hat{d}^\dag_\uparrow \hat{d}_\uparrow \hat{d}^\dag_\downarrow \hat{d}_\downarrow + \hat{H}_\textup{leads}+\hat{H}_\textup{tun},
\end{equation}
where the impurity instead has onsite Coulomb repulsion with interaction strength $U_d$. 
In the AIM, a strong onsite repulsion generates correlations by imposing a constraint of no double occupancy on the impurity site. In our model \eqref{eq: master equation}, we achieve the same constraint by introducing a dissipation that actively removes double occupancies. However, we stress that in our model there are no interactions; all the correlations are introduced by two-body losses.

\section{Method}\label{sec: method}

In this section, we introduce and motivate the SGS variational ansatz in the context of closed systems.
We then extend this method to open-system impurity models by using the quantum trajectory approach.
The main achievement of our work is the combination of SGS with quantum trajectory, whose analytical details are presented in Sec. \ref{sec: SGS techniques}.  

\subsection{Superposition of Gaussian states ansatz}

We begin by introducing the SGS ansatz for closed systems. Considering the AIM with Hamiltonian \eqref{eq: AIM}, the hybridization between the impurity and the bath leads to charge fluctuations at the impurity site. 
A general wave function can be expressed as a superposition of four states that feature different configurations in the impurity subspace, namely
\begin{equation} \label{eq: a general state}
    \ket{\Psi}=\sum_{s=0, \uparrow, \downarrow, 2} \alpha_s \ket{s}\ket{\psi_s}.
\end{equation}
Here, $\ket{s}$ denotes the impurity states $\ket{0}$ (vacuum state), $\ket{\uparrow}$, $\ket{\downarrow}$, or $\ket{2}$ (double occupation), and $\ket{\psi_s}$ represents the corresponding many-body state of the bath. The coefficients $\alpha_s$ capture the entanglement between the impurity and the bath.

To gain intuition about the structure of the bath states $\ket{\psi_s}$, we first consider the case when the tunneling $\hat{H}_\rm{tun}$ is turned off  and the bath is in its ground state, the Fermi sea $\ket{\rm{FS}}$. When the tunneling is turned on, the bath becomes connected to the impurity. As $\hat{H}_\rm{tun}$ transfers electrons into and out of the bath, it generates both charge fluctuations and particle-hole excitations in the bath. 
This idea is the basis of the Gunnarson-Sch\"onhammer ansatz, which was successfully employed to predict both the ground-state properties and the spectral function of the AIM \cite{Gunnarsson1983}. 
A simple way to further improve the ansatz is to assume Gaussian states for the baths within different impurity subsectors, namely an ansatz using a superposition of four Gaussian states:
\begin{equation} \label{eq: SGS ansatz}
    \ket{\Psi_{\rm{SGS}}}=\sum_{s=0, \uparrow, \downarrow, 2} \alpha_s\ket{s}\ket{\rm{GS}_s},
\end{equation}
which increases the number of variational parameters while remaining numerically efficient. Here, the Gaussian state $\ket{\rm{GS_s}}$ corresponds to a single Slater determinant and can capture the particle-hole excitations non-perturbatively, which will be precisely defined in Sec.~\ref{sec: SGS techniques}.
To describe closed-system dynamics, we consider $\alpha$ and $\ket{\rm{GS}_s}$ to be time-dependent. 
The technical details of implementing the real-time variational SGS ansatz are also discussed in Sec.~\ref{sec: SGS techniques}.

It should be noted that Bravyi and Gosset \cite{BravyiGosset_complexity} have shown that the ground state of an impurity Hamiltonian, up to a small fixed error, can be approximated as a superposition of a few Gaussian states, with the number of states scaling polynomially with the size of the bath.
The SGS ansatz \eqref{eq: SGS ansatz} used in our work can be considered as a ``minimal version" of the most general SGS and thus  highly efficient.
The variational SGS has been implemented in closed systems and has been shown to capture both ground-state \cite{boutin2021quantum} and dynamical properties \cite{nondissipativeSGS}.

The SGS ansatz works well with pure states, both in and out of equilibrium.
For our dissipative problem---and more generally---we aim to further extend it to open-system impurity problems. The simplest way is to employ the quantum trajectory approach, which allows approximating a mixed state with an ensemble of pure states. This enables the direct use Eq.~\eqref{eq: SGS ansatz}.
To explore the dynamics governed by the master equation \eqref{eq: master equation}, we focus exclusively on real-time evolution, starting from a pure state of the entire system. 

\subsection{Quantum trajectory approach\label{sec: quantum trajectory approach}}
In this subsection, to make our work self-contained, we briefly revisit the quantum trajectory approach, previously detailed in the literature \cite{Dalibard1992,Molmer1993,Dum1992,Dum1992_2,carmichael1993open}. 

Considering the master equation \eqref{eq: master equation}, we rewrite it as
\begin{equation} \label{eq: general master equation}
    \dt \hat{\rho} = - i (\hat{H}_\rm{NH}\,\hat{\rho} - \hat{\rho}\,\hat{H}_\rm{NH}^\dag) + \gamma \hat{L}^\dag \rho \hat{L},
\end{equation}
with a non-Hermitian Hamiltonian
\begin{equation} \label{eq: non-Hermitian Ham}
   \hat{ H}_\rm{NH} = \hat{H}-i\frac{\gamma}{2}\hat{L}^\dag \hat{L}.
\end{equation}
It has been proven that Eq.~\eqref{eq: general master equation} can be reproduced by the following unraveling procedure. For simplicity, we assume that the initial density operator $\hat{\rho}(t=0)=\ketbra{\psi_0}$ is a pure state, with $\ket{\psi_0}$ denoting the normalized initial wave function. Then, one forms an ensemble of $N_\rm{Traj}$ quantum trajectories $\ket{\psi_i(t)}$ by evolving $\ket{\psi_0}$ with $\hat{H}_\rm{NH}$, and interspersing the evolution with discontinuous quantum jumps $\ket{\psi_i(t^+)}=\hat{L}\ket{\psi_i(t^-)}/\norm*{\hat{L}\ket{\psi_i(t^-)}}$ according to certain probabilistic rules. The average state of the ensemble thus constructed, 
\begin{equation} \label{eq: density operator with QT}
    \hat{\rho}(t)=\frac{1}{N_\rm{Traj}}\sum_{i=1}^{N_\rm{Traj}} \frac{\ketbra{\psi_i(t)}}{\braket{\psi_i(t)}{\psi_i(t)}},
\end{equation}
converges to the solution of Eq.~\eqref{eq: general master equation} as $N_\rm{Traj} \rightarrow \infty$. In a concrete simulation with a finite $N_\rm{Traj}$, Eq.~\eqref{eq: density operator with QT} provides an approximation to the desired $\hat{\rho}(t)$.

Following Ref.~\cite{DaleyQuantumTrajectories}, we outline the main procedures on how to obtain $\ket{\psi_i(t)}$:
(a) Draw a random number $r$ sampled from a uniform distribution over the interval $[0,1]$. 
(b) Solve the non-Hermitian real-time evolution $\ket{\psi_i(t)}=\rm{exp}\,(- i  \hat{H}_\rm{NH}t)\ket{\psi_0}$ with the initial state $\ket{\psi_0}$ and the non-Hermitian Hamiltonian.
(c) Solve equation $\norm{\ket{\psi_i(t)}}^2=r$ numerically to find the time $t_1$ when the quantum jump occurs. 
(d) At time $t_1$, apply the jump and then normalize the new wave function as $\ket{\psi_i(t_1^+)}=\hat{L}\ket{\psi_i(t_1^-)}/\norm*{\hat{L}\ket{\psi_i(t_1^-)}}$. 
(e) Continue the time evolution from step (a) by sampling a new random number $r$ and replacing the initial state with $\ket{\psi_i(t_1^+)}$. 

The expectation value of an observable $\hat{O}$ is obtained by taking the average over the contributions from all trajectories, namely
\begin{equation}
    O(t)\equiv\rm{Tr}(\hat{O}\hat{\rho}(t))=\frac{1}{N_\rm{Traj}}\sum_{i=1}^{N_\rm{Traj}} O_i(t),
\end{equation}
with $O_i(t) \equiv \bra{\psi_i(t)}\hat{O}\ket{\psi_i(t)}/\braket{\psi_i(t)}{\psi_i(t)}$ being the expectation value for a single trajectory  $i$.
For large but finite $N_\rm{Traj}$, the statistical error of $O(t)$
\begin{equation}
    \sigma_O=\frac{\Delta O}{\sqrt{N_\rm{Traj}}} 
\end{equation}
asymptotically approaches zero, as $\sigma_O\sim N_\rm{Traj}^{-1/2}$. Here, $\Delta O=\sum_i(O_i-O)^2/N_\rm{Traj}$ is the standard deviation, which generally converges as $N_\rm{Traj}\rightarrow\infty$. 

The quantum trajectory approach has the advantage of evolving pure states of dimension $N_\mathcal{H}$ instead of the density operator of dimension $N_\mathcal{H}^2$, where the Hilbert space dimension $N_\mathcal{H}$ generally increases exponentially with the size of the system. Therefore, it is very efficient for solving open-system quantum problems. The trade-off is the introduction of statistical errors resulting from the (classical) fluctuations in the sampling of a finite number of trajectories.
Nevertheless,  there is no sign problem in this approach (each trajectory carries equal and positive weight), so the errors can be well controlled by increasing the number of trajectories. 

\subsection{Variational techniques\label{sec: SGS techniques}}
In this subsection, we provide the technical details of implementing our open-quantum-system SGS approach. 
The SGS ansatz~\eqref{eq: SGS ansatz} consists of a superposition of four direct product states $\ket{s}\ket{\rm{GS}_s}$ in different impurity sectors.
Between two successive quantum jumps, the non-Hermitian Hamiltonian \eqref{eq: non-Hermitian Ham} for describing the dynamics explicitly reads $\hat{H}_\rm{NH}=\hat{H}-i\frac{\gamma}{2}\hat{n}_{d\uparrow} \hat{n}_{d\downarrow}$, which is exactly the AIM (see Eq.~\eqref{eq: AIM}) with an imaginary interaction. 
The total fermion number $\hat{N}_\rm{tot}=\sum_\sigma \hat{d}^\dag_\sigma \hat{d}_\sigma+\sum_{j\sigma\alpha} \hat{c}_{j\sigma\alpha}^\dagger \hat{c}_{j\sigma\alpha}$ is conserved, since it commutes with $H_\rm{NH}$. The Gaussian state in the SGS ansatz is defined by
\begin{equation} \label{eq: Gaussian state}
    \ket{\rm{GS}_s(t)} \equiv e^{i \hat{C}^{\dagger} \xi_s(t) \hat{C}}\ket{\rm{Fock}_s},
\end{equation}
which has a definite particle number $N_s=N_\rm{tot}, N_\rm{tot}-1, N_\rm{tot}-2$ for $s=0,\uparrow(\downarrow),2$, respectively. Denoting by $N_f$ the number of bath modes, the variational parameter $\xi_s$ is a $N_f \times N_f$ Hermitian matrix  which means the SGS ansatz \eqref{eq: SGS ansatz} has $O(N_f^2)$ variational parameters. The column vector $\hat{C}=(\hat{c}_{1},\hat{c}_2,\dots,\hat{c}_{N_f})^\rm{T}$ is a shorthand for the ordered combination of all annihilation operators $\hat{c}_{p \sigma \alpha}$ in the bath. Without loss of generality, the Fock state $\ket{\rm{Fock}_s}$ is fixed to $\hat{C}_1^\dagger \hat{C}_2^\dagger...\hat{C}_{N_s}^\dagger\ket{0_\rm{bath}}$, in which the first $N_s$ modes on top of the vacuum state $\ket{0_\rm{bath}}$ of the leads are occupied.
For clarity, we note that the Gaussian state \eqref{eq: Gaussian state} with a definite particle number can be put in the standard Slater determinant form as $\ket{\rm{GS}_s}=\hat{\bar{C}}_1^\dagger \hat{\bar{C}}_2^\dagger \cdots \hat{\bar{C}}_{N_s}^\dagger \ket{0}$ by letting $e^{i \hat{C}^\dag \xi_s \hat{C}}$ transform the bath modes $\hat{C}$ to $\hat{\bar{C}}^\dagger = \hat{C}^\dagger U_s$ with $U_s = \exp(i \xi_s)$.

To employ the quantum trajectory approach, we need two ingredients: how to solve the non-Hermitian evolution between quantum jumps and how to implement these quantum jumps. Let us start with the non-Hermitian evolution.
Through the time-dependent variational principle (TDVP) \cite{haegeman2011,haegeman2016}, the variational dynamics can be determined by minimizing the variational error $\norm*{\dt\ket{\Psi_{\rm{SGS}}} + i \hat{H}_\rm{NH}\ket{\Psi_{\rm{SGS}} }}$ at each time, where $\norm{\dots}$ denotes the standard Hilbert space norm.
This minimization leads to the (non-Hermitian) projective Schr\"odinger equation
\begin{equation} \label{eq: real-time evolution}
    \dt\left|\Psi_{\rm{SGS}}\right\rangle=-i \hat{P}_{\mathcal{T}} \hat{H}_\rm{NH} \left|\Psi_{\rm{SGS}}\right\rangle.
\end{equation}
Here, $\hat{P}_{\mathcal{T}}$ projects any state into the tangent space of the variational manifold, which has orthonormal basis vectors $\ket{s}\ket{\rm{GS}_s}$ and $ e^{i \hat{C}^{\dagger} \xi_s \hat{C}}\hat{C}^\dagger_j \hat{C}_{j'}\ket{s}\ket{\rm{Fock}_s}$ with $j \in \mathcal{I}_\mathcal{P},j'\in \mathcal{I}_\mathcal{H}$.
The indices $\mathcal{I}_\mathcal{P}=\{N_s+1,\dots,N_f\}$ correspond to the created particles (indexed by $j$) on top of $\ket{\rm{Fock}_s}$. Analogously, the indices $\mathcal{I}_\mathcal{H}=\{1,\dots,N_s\}$ correspond to the generated holes (indexed by $j'$).

Choosing $\alpha_s$ and $U_s=\exp(i\xi_s)$ as time-dependent variational parameters,
the left-hand side of Eq.~\eqref{eq: real-time evolution} can be expressed as \cite{shi2018}
\begin{equation} \label{eq: LHS of real-time evolution}
    \dt \left|\Psi_\rm{SGS}\right\rangle=\sum_s e^{i\hat{C}^{\dagger} \xi_s \hat{C}}\left[\dot{\alpha}_s+\alpha_s \hat{C}^{\dagger} U_s^{\dagger} \dot{U}_s \hat{C}\right]|s\rangle|\rm{Fock}_s\rangle.
\end{equation}
In the formula above, we notice that $\hat{c}_{j_1}^\dagger \hat{c}_{j_2}|\rm{Fock}_s\rangle = 0$ for $j_2 \in \mathcal{I}_\mathcal{P}$  gives a trivial contribution and $\hat{c}_{j_1}^\dagger \hat{c}_{j_2}|\rm{Fock}_s\rangle = \delta_{j_1 j_2}|\rm{Fock}_s\rangle$ for $j_1,j_2 \in \mathcal{I}_\mathcal{H}$ gives the same contribution as the $\dot{\alpha}_s$ term. This means that there are redundancies in the variational parameters, namely multiple sets of $\dot{\alpha}_s$ and $\dot{U}_s$ that give the same physical state on the right-hand side of Eq.~\eqref{eq: real-time evolution}. To remove these redundancies, it is sufficient to apply the constraints
\begin{equation}\label{eq: fix redundencies}
\begin{aligned}
        (U_s^{\dagger} \dot{U}_s)_{\mathcal{I}_\mathcal{H} \mathcal{I}_\mathcal{H}}&=0_{N_s\cross N_s }, \\
    (U_s^{\dagger} \dot{U}_s)_{\mathcal{I}_\mathcal{P} \mathcal{I}_\mathcal{P}}&=0_{(N_f-N_s)\cross (N_f-N_s) }.
\end{aligned}
\end{equation}
These constraints are applied to the hole and particle subblocks in matrix $U_s^{\dagger} \dot{U}_s$.

The equations of motion (EoMs) for $\dot{\alpha}_s$ and $\dot{U}_s$ are determined by taking the overlap between $-i \hat{H}_\rm{NH} \ket{\Psi_\rm{SGS}}$ and the tangential basis:
\begin{equation} \label{eq: overlap}
    \begin{aligned}
    \dot{\alpha}_s &=-i{\bra{\rm{GS}_s}}\bra{s} \hat{H}_\rm{NH} \ket{\Psi_{\rm{SGS}}}, \\
    (U_s^{\dagger} \dot{U}_s)_{j j'} & = -i \alpha_s^{-1}  \bra{\rm{GS}_s}\bra{s} \hat{c}_{j'}^\dagger \hat{c}_j  \hat{H}_\rm{NH} \ket{\Psi_{\rm{SGS}}}
\end{aligned}
\end{equation}
with $j \in \mathcal{I}_\mathcal{P}$ and $j'\in \mathcal{I}_\mathcal{H}$. The other elements are fully fixed by the constraints in Eq.~\eqref{eq: fix redundencies} and the unitarity of $U_s$, which gives $(U_s^{\dagger}\dot{U}_s)^\dagger = -U_s^{\dagger} \dot{U}_s$. 
To evaluate the right-hand side of Eq.~\eqref{eq: overlap}, we need to calculate the action of $\hat{H}_\rm{NH}$, which can be recast as $\hat{H}_\rm{NH}=\hat{\bar{H}}_\rm{imp}+\hat{H}_\rm{leads}+\hat{H}_\rm{tun}$ with $\hat{\bar{H}}_\rm{imp}=\hat{H}_\rm{imp}-i\frac{\gamma}{2}\hat{n}_{d\uparrow} \hat{n}_{d\downarrow}$.  The action of $\hat{\bar{H}}_\rm{imp}$ is straightforward as it acts only on the impurity site. The contribution of $\hat{H}_\rm{leads}$ can easily be obtained utilizing the Wick contraction theorem for the individual Gaussian state $\ket{\rm{GS}_s}$. The action of $\hat{H}_\rm{tun}$ is the only action that results in a variational error and is nontrivial because it has matrix elements between different impurity states. Meanwhile, it generates new Gaussian states through a single annihilation (creation) operator $\hat{c}_{\sigma}'$ ($\hat{c}'^\dagger_{\sigma}$) with $\hat{c}_{\sigma}' = \sum_{p\alpha}V_{p\alpha}\hat{c}_{p\sigma\alpha}$, namely
\begin{equation} \label{eq: new GS}
\begin{aligned}
    \hat{c}_{\sigma}'\ket{\rm{GS}_s} = \beta^-_{\sigma s}\ket{\rm{GS}_{\sigma s-}}, \\
    \hat{c}_{\sigma}'^\dagger\ket{\rm{GS}_s} = \beta^+_{\sigma s}\ket{\rm{GS}_{\sigma s +}}.
\end{aligned}
\end{equation}
Here,  the generated state $\ket{\rm{GS}_{\sigma s \mp}}$ and the relevant coefficient $\beta^\mp_{\sigma s}$ are obtained by manipulating the occupied modes of $\ket{\rm{GS}_s}$ (see Appendix \ref{appendix: manipulation of GS} for details). 
Finally, we need to calculate the overlaps between the different Gaussian states,  which can be derived through the following expressions \cite{Zhang1997,snyman2021}:
\begin{subequations} \label{eq: overlaps expressions}
\begin{align}
    \mathcal{O}_{\lambda,\lambda'} &= \braket{\rm{GS}_\lambda}{\rm{GS}_{\lambda'}} = \rm{det}( u_\lambda^\dag  u_{\lambda'}), \\    
    (\mathcal{\varrho}_{\lambda, \lambda'})_{ij} &= \bra{\rm{GS}_\lambda} 
    \hat{C}_i^{\dagger} \hat{C}_j\ket{\rm{GS}_{\lambda'}}    \notag\\    &=\mathcal{O}_{\lambda,\lambda'}\left[u_{\lambda'}\left(u_\lambda^\dag u_{\lambda'}\right)^{-1} u_\lambda^{\dagger}\right]_{ji}   \notag\\
    &=\left[u_{\lambda'}\,\rm{adj}(u_\lambda^\dag u_{\lambda'})u_\lambda^{\dagger} \right]_{ji}\label{eq: rho12},
\end{align}
\end{subequations}
where $\ket{\rm{GS}_\lambda}$ and $\ket{\rm{GS}_{\lambda'}}$, having the same particle number $\mathcal{N}_\lambda$, denote the two different Gaussian states. The $N_f\times \mathcal{N}_\lambda$ matrix $u_\lambda$ ($u_{\lambda'}$), representing the first $\mathcal{N}_\lambda$ columns of $U_\lambda$ ($U_{\lambda'}$), consists of all orthonormal modes occupied in the state $\ket{\rm{GS_\lambda}}$ ($\ket{\rm{GS_{\lambda'}}}$).
The last line of Eq.~\eqref{eq: rho12} can be applied even if the matrix $u_\lambda^\dag u_{\lambda'}$ is not invertible, where $\rm{adj}(\dots)$ represents the adjugate of a square matrix.

Following the steps outlined above and using equations \eqref{eq: new GS} and \eqref{eq: overlaps expressions}, 
 we finally obtain the EoMs \eqref{eq: overlap} for the variational parameters between two quantum jumps:
\begin{subequations}
\begin{align}
    \dot{\alpha}_s &= -i\left(n_s\epsilon_d+E_s-i \frac{\gamma}{2} \delta_{s 2}\right) \alpha_s+f_s, \\
    \dot{U}_s &= -i U_s \mathcal{A}_s,    
\end{align}
\end{subequations}
where $n_s\equiv \bra{s}\hat{n}_{d\uparrow}+\hat{n}_{d\downarrow}\ket{s}$ is the particle number of the impurity and
\begin{subequations}
\begin{align}
E_s &= \left\langle\rm{GS}_s\left|H_{\text {leads }}\right | \rm{GS}_s\right\rangle = \rm{Tr} \left(U_s^{\dagger} h U_s\right)_{\mathcal{HH}}, \\
\mathcal{A}_s &= \left[\begin{array}{cc}
0_{N_s\cross N_s } & \left[U_s^{\dagger} (h+\mathcal{F}_s) U_s\right]_{\mathcal{HP}} \\
\left[U_s^{\dagger} (h+\mathcal{F}_s) U_s\right]_{\mathcal{PH}} & 0_{(N_f-N_s)\cross (N_f-N_s) }
\end{array}\right].
\end{align}    
\end{subequations}
Here, the Hermitian matrix $h$ is defined by rewriting $\hat{H}_{\rm {leads}}=\hat{C}^\dagger h \hat{C}$.
The variables $f_s$ and $\mathcal{F}_s$ originate from the tunneling events, and are summarized below:
\begin{equation}
    \begin{split}    
f_0= \alpha_{\uparrow} \beta_{\uparrow \uparrow}^{+} \mathcal{O}_{0, \uparrow \uparrow+}+\alpha_{\downarrow} \beta_{\downarrow \downarrow}^{+} \mathcal{O}_{0, \downarrow \downarrow+}, \\
f_{\uparrow}= \alpha_0 \beta_{\uparrow 0}^{-}\mathcal{O}_{\uparrow, \uparrow 0-}+\alpha_2 \beta_{\downarrow 2}^{+}\mathcal{O}_{\uparrow, \downarrow 2+}, \\
f_{\downarrow}= \alpha_0 \beta_{\downarrow 0}^{-}\mathcal{O}_{\downarrow, \downarrow 0-} - \alpha_2 \beta_{\uparrow 2}^{+} \mathcal{O}_{\downarrow, \uparrow 2+},\\
f_2= \alpha_{\uparrow} \beta_{\downarrow \uparrow}^{-} \mathcal{O}_{2, \downarrow \uparrow-}-\alpha_{\downarrow} \beta_{\uparrow \downarrow}^{-}\mathcal{O}_{2, \uparrow \downarrow-},
    \end{split}
\end{equation}
and
\begin{equation} \label{eq: F variables}
    \begin{split}
        \begin{aligned}
& \mathcal{F}_0= \alpha_0^{-1}\left(\alpha_{\uparrow} \beta_{\uparrow \uparrow}^{+} \varrho_{0, \uparrow \uparrow+}+\alpha_{\downarrow} \beta_{\downarrow \downarrow}^{+} \varrho_{0, \downarrow \downarrow+}\right), \\
& \mathcal{F}_{\uparrow}= \alpha_{\uparrow}^{-1}\left(\alpha_0 \beta_{\uparrow 0}^{-} \varrho_{\uparrow, \uparrow 0-}+\alpha_2 \beta_{\downarrow 2}^{+} \varrho_{\uparrow, \downarrow 2+}\right), \\
& \mathcal{F}_{\downarrow}= \alpha_{\downarrow}^{-1}\left(\alpha_0 \beta_{\downarrow 0}^{-} \varrho_{\downarrow, \downarrow 0-} - \alpha_2 \beta_{\uparrow 2}^{+} \varrho_{\downarrow, \uparrow 2+}\right), \\
& \mathcal{F}_2= \alpha_2^{-1}\left(\alpha_{\uparrow} \beta_{\downarrow \uparrow}^{-} \varrho_{2, \downarrow \uparrow-}-\alpha_{\downarrow} \beta_{\uparrow \downarrow}^{-} \varrho_{2, \uparrow \downarrow-}\right),
\end{aligned}
    \end{split}
\end{equation}
where we have introduced the notation $\mathcal{O}_{s,\lambda}$ and $\varrho_{s,\lambda}$, with $s$ and $\lambda$ denoting the index of $\ket{\rm{GS}_{s}}$ and $\ket{\rm{GS}_{\sigma s' \pm}}$, respectively.
The quantities $\mathcal{O}_{s,\lambda}$ and $\varrho_{s,\lambda}$ are obtained through Eqs. \eqref{eq: new GS} and \eqref{eq: overlaps expressions} along with the manipulations detailed in Appendix \ref{appendix: manipulation of GS}.

The only missing ingredient is how to implement quantum jumps that occur randomly during evolution. 
Due to the local nature of the jump operator $L$, the effect of quantum jumps on the SGS is simple: every impurity state will be annihilated, except for $\ket{2}$, which is then mapped to $\ket{0}$.
Correspondingly, assuming the quantum jump occurs at $t_1$, the variational parameters should be updated by
\begin{equation}\label{eq: quantum jump}
    \begin{aligned}
        \alpha_0(t_1^+)&=\alpha_2(t_1^-), \, \alpha_{\uparrow,\downarrow,2}(t_1^+)=0, \\
         U_0(t_1^+)&=U_2(t_1^-).
    \end{aligned}
\end{equation}
To determine $U_{\uparrow,\downarrow,2}(t_1^+)$, we consider the evolution over a short time interval after the jump, utilizing the Taylor expansion of the non-Hermitian evolution (see Appendix~\ref{appendix: evolution after jump}). 

In summary, the SGS techniques, with $O(N_f^2)$ variational parameters, can efficiently implement non-Hermitian evolution and quantum jumps in the quantum trajectory approach. This variational method, as shown in the following sections, is non-perturbative and works well in different regimes of loss rate $\gamma$.
Our method can also be directly generalized to other impurity problems in an open system, with a superconducting bath; a superfluid bosonic bath; or multiple impurity sites (or quantum dots), etc.

\section{Results: crossover to Kondo regime} \label{sec: results}
In this section, we apply our SGS approach to a specific implementation of the model outlined in Sec.~\ref{sec: model}: we consider two reservoirs modeled as two chains of nearest-neighbor hopping fermions with open boundary conditions, connected to the dot site at one end:
\begin{subequations} \label{eq: lattice model}
    \begin{align}
        H_\textup{leads}=&-J\sum_{\alpha=L,R}\sum_{j=1}^{\ell-1}\sum_{\sigma}(\hat{c}_{j+1\sigma\alpha}^\dag \hat{c}_{j\sigma\alpha}+\rm{H.c.})\notag\\
        &-\sum_{\alpha=L,R}\sum_{j=1}^{\ell}\sum_{\sigma}\mu_\alpha \hat{c}_{j\sigma\alpha}^\dag \hat{c}_{j\sigma\alpha} \label{Ham: leads}\\
        H_\textup{tun}=&-V\sum_{\alpha=L,R}\sum_{\sigma}(\hat{d}_{\sigma}^\dag \hat{c}_{1\sigma\alpha}+\rm{H.c.}),\label{Ham: tun}
    \end{align}
\end{subequations}
where $\ell$ is the size of each chain, $J$ is the hopping amplitude of the fermions in the bath, and $V$ is the tunneling amplitude. The channel index $\alpha=L,R$ represents the left and right reservoirs.
This choice of the leads is not particularly restrictive, since impurity problems can always be reduced to a collection of one-dimensional (1D) channels \cite{GogolinNersesyanTsvelik,Giamarchi}. The leads' Hamiltonian can be diagonalized by $\hat{c}_{j\sigma\alpha}=\sum_{p}\psi_p(j) \hat{c}_{p\sigma\alpha}$, with $\psi_p(j)=\sqrt{\frac{2}{\ell+1}}\sin pj$ where $p\in\{\frac{\pi}{\ell+1}n| \, n=1...\ell\}$. The resulting dispersion and impurity-lead tunneling are
\begin{subequations}\label{eq: leads properties}
\begin{align}
    \ce_{p\alpha}&=-2J\cos p - \mu_\alpha,\\
    V_{p\alpha}&=-V \psi_p(1)=-\sqrt{\frac{2}{\ell+1}} V \sin p.
\end{align}
\end{subequations}
Through Fermi's golden rule, the tunneling rate associated with the $\alpha$ channel is given by
\begin{equation} \label{eq: tunneling rates}
\begin{aligned}
        \Gamma_\alpha(\omega)&= 2\pi\sum_p |V_{p\alpha}|^2 \delta(\omega-\ce_{p\alpha}) \\
       & =\frac{2V^2}{J}\sqrt{1-\frac{(\omega-\mu_\alpha)^2}{4J^2}}
\end{aligned}
\end{equation}
for $|\omega-\mu_\alpha|\leq 2J$ and zero otherwise. The total tunneling rate of the impurity is then given by $\Gamma(\omega)=\Gamma_L(\omega)+\Gamma_R(\omega)$.

We employ the SGS approach to show how the Kondo effect emerges as a consequence of the localized two-body losses by characterizing two signatures of Kondo physics: spin relaxation and charge conductance. In both cases, the non-perturbative nature of the SGS ansatz allows us to examine the full crossover from small to large $\gamma$.

Our simulations are performed using $N_{\rm{Traj}}=1000$ trajectories, ensuring statistical reliability. With $95\%$ confidence, the statistical error in our Monte Carlo simulations is generally below $2\%$, and this applies to all quantum trajectory-based figures in this work.

\subsection{Benchmark of the variational approach} \label{sec: benchmark}

We begin by applying the SGS approach to a small system (with the size of each chain being $\ell=5$), considering the loss scenario where no local Hubbard interaction is present, and comparing the results under varying dissipation rates $\gamma$ with the ED results. We fix the chemical potentials and the onsite potential of the impurity to $\mu_L=\mu_R=\ce_d=0$, so that the system is at half filling.  The tunneling amplitude is set to $V=J$. It should be noted that the system exhibits parity symmetry, characterized by the interchangeability of the channels $L$ and $R$. Therefore, the impurity will only couple to the symmetric bath modes $\hat{c}_{i\sigma e}=(\hat{c}_{i\sigma L}+\hat{c}_{i\sigma R})/\sqrt{2}$, while the asymmetric bath modes $\hat{c}_{i \sigma o}=(\hat{c}_{i\sigma L}-\hat{c}_{i \sigma R})/\sqrt{2}$  are decoupled. The two-channel model \eqref{eq: lattice model} is then mapped to a one-channel model with the tunneling amplitude replaced by $\sqrt{2}V$. In both ED (performed with the QuTiP package \cite{qutip1,qutip2}) and SGS calculations, we only consider $6$ symmetric sites, including the impurity site.

\begin{figure}[t]
    \centering
    \includegraphics[width=1\linewidth]{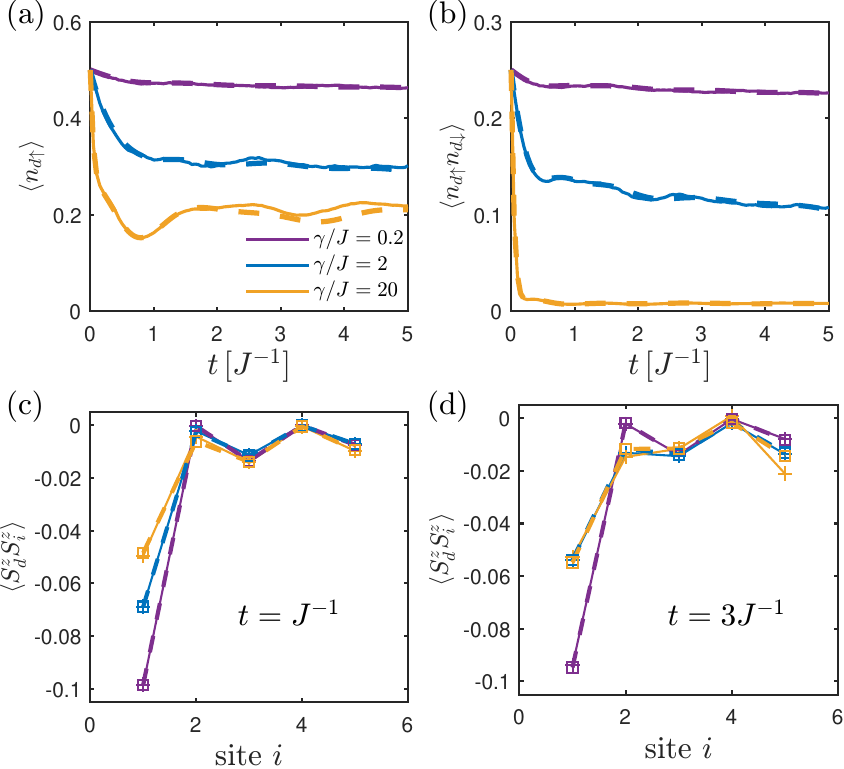}
\caption{Benchmark between the SGS approach (solid curves) and the ED approach (dashed curves) in a small system of $\ell=5$. We consider only the symmetric modes. The dissipation is first turned off and the system is prepared in the ground state with the numbers of spin-up and spin-down particles being $N_{\mathrm{tot},\uparrow}=N_{\mathrm{tot},\downarrow}=3$. Then the dissipation is quenched to a finite value, and the state starts evolving.  (a) and (b) show the time-dependent $\langle \hat{n}_{d\uparrow} \rangle$ and $\langle \hat{n}_{d\uparrow}\hat{n}_{d\downarrow}\rangle$, in which the quantum Zeno effect occurs for $\langle \hat{n}_{d\uparrow}\hat{n}_{d\downarrow}\rangle$ at $\gamma/J=20$. (c) and (d) show spin-$z$ correlation between the impurity and bath sites at different times $t=J^{-1}$ and $t=3J^{-1}$, which exhibit antiferromagnetic signatures. The parameters used are $V=J, \mu_L=\mu_R=\ce_d=0$, and $N_\rm{Traj}=1000$. With $95\%$ confidence, the statistical error in our Monte Carlo simulations is generally below $2\%$, and this applies to all quantum trajectory-based figures in this work.
}
    \label{fig:benchmark1}
\end{figure}

To simulate the dynamics, we first turn off the dissipation and prepare the whole system in the ground state, with the total numbers of the spin-up and spin-down particles being $N_{\mathrm{tot},\uparrow}=N_{\mathrm{tot},\downarrow}=3$. We then quench the dissipation to a finite value and let the state evolve under the master equation~\eqref{eq: master equation}.
Due to the spin symmetry of both the Hamiltonian and the initial (ground) state, the spin occupations remain equal throughout the evolution, i.e., $ \langle \hat{n}_{d\uparrow}\rangle=\langle\hat{n}_{d\downarrow}\rangle$.
In Figs.~\ref{fig:benchmark1}(a) and \ref{fig:benchmark1}(b), we show the time-dependent expectation values of the spin-up particle occupation $\langle \hat{n}_{d\uparrow} \rangle$ and the double occupation $ \langle \hat{n}_{d\uparrow}\hat{n}_{d\downarrow}\rangle$, which can be directly obtained from our SGS ansatz~\eqref{eq: SGS ansatz}.
These physical quantities capture how two-body dissipation affects particle occupations on the impurity site.
For a low dissipation rate $\gamma=0.2 J$ (blue curves), $\langle \hat{n}_{d\uparrow} \rangle$ and $\langle \hat{n}_{d\uparrow} \hat{n}_{d\downarrow}\rangle$ are almost unchanged during the evolution, as the system is close to the non-dissipative case of $\gamma=0$, whose ground state can be exactly expressed by the SGS ansatz.
For an intermediate dissipation rate $\gamma=2J$ (orange curves), both $\langle \hat{n}_{d\uparrow} \hat{n}_{d\downarrow}\rangle$ and $\langle \hat{n}_{d\uparrow} \rangle$ display a two-step dynamics, with a steeper transient followed by a slower decay. We find the transient in the double occupancies occurs with a rate $\sim\gamma$, while the spin-up occupation decays more slowly with a rate $\sim\gamma/2$.
This difference in timescales between $\langle \hat{n}_{d\uparrow} \rangle$ and $\langle \hat{n}_{d\uparrow} \hat{n}_{d\downarrow}\rangle$ can be understood by considering that weak two-body losses do not impart strong correlations between spin-up and spin-down sectors, so we can estimate $\langle \hat{n}_{d\uparrow}\hat{n}_{d\downarrow}\rangle\simeq \langle \hat{n}_{d\uparrow} \rangle \langle \hat{n}_{d\downarrow} \rangle=\langle \hat{n}_{d\uparrow} \rangle^2$.
Therefore, the decay rate of $\langle \hat{n}_{d\uparrow} \hat{n}_{d\downarrow}\rangle$ is about twice that of $\langle \hat{n}_{d\uparrow} \rangle$, namely $\dt \langle \hat{n}_{d\uparrow} \hat{n}_{d\downarrow}\rangle / \langle \hat{n}_{d\uparrow} \hat{n}_{d\downarrow}\rangle\simeq 2\dt \langle \hat{n}_{d\uparrow}\rangle / \langle \hat{n}_{d\uparrow}\rangle $.
For a high dissipation rate $\gamma=20 J$ (purple curves), $\langle \hat{n}_{d\uparrow} \hat{n}_{d\downarrow}\rangle$ drops dramatically at first and then remains nearly zero, which is a clear signature of the quantum Zeno effect due to the fast decay of the double occupation state $\ket{\uparrow\downarrow}$ on the impurity site. However, the dynamics of $\langle \hat{n}_{d\uparrow} \rangle$ has two stages corresponding to different energy scales: the first stage is the fast decay with a rate of $\gamma$ within the initial short time interval $t\lesssim \gamma^{-1}$, resulting from the rapid two-body decay of the impurity state $\ket{\uparrow\downarrow}$. The second stage is a slower decrease followed by slight fluctuations around an average value, which results from charge tunneling and fluctuations within the impurity sectors $\ket{0}$ and $\ket{\uparrow}$.
For all regimes of $\gamma$, the SGS results agree well with the ED results, although $\langle \hat{n}_{d\uparrow}\rangle$ shows slightly larger discrepancies for $\gamma=20J$ after a long time evolution when $t\gtrsim 2.5 J^{-1}$.

To further testify the accuracy of our approach, we investigated the spin-$z$ correlation functions between the impurity site and the bath sites at two distinct times, $t=J ^{-1}$ and $t=3 J ^{-1}$. These correlations are denoted as $\langle \hat{S}_d^z \hat{S}_{i}^z\rangle$, in which $\hat{S}_d^a=\frac{1}{2}\sum_{\sigma\sigma'}\hat{d}_\sigma^\dag\tau^a_{\sigma\sigma'}\hat{d}_{\sigma'}$ and $\hat{S}_{i}^a=\frac{1}{2}\sum_{\sigma\sigma'}\hat{c}^\dag_{i\sigma e}\tau^a_{\sigma\sigma'}\hat{c}_{i\sigma' e}$ (with the Pauli matrices $\tau^a$) represent the spin operators of the impurity and the symmetric bath site $i$, respectively. As presented in Figs.~\ref{fig:benchmark1}(c) and \ref{fig:benchmark1}(d),  $\langle \hat{S}_d^z \hat{S}_{i}^z\rangle$ is negative, indicating antiferromagnetic (AF) signatures for different dissipation rates, especially for odd bath sites. At $t=J^{-1}$,  the magnitude of $\langle \hat{S}_d^z \hat{S}_{1}^z\rangle$ decreases with increasing $\gamma$, showing that the two-body loss on the impurity site tends to suppress the AF correlations present in the ground state with $\gamma=0$.  This phenomenon can be explained as follows \cite{Nakagawa2020}.  For simplicity, consider only the impurity site and the first bath site, initialized in an AF state $\ket{\uparrow}_\rm{imp}\ket{\downarrow}_1$. This state can be transferred to the double occupation state $\ket{\uparrow\downarrow}_\rm{imp}\ket{0}_1$ through tunneling $\hat{H}_\mathrm{tun}$, thus it will undergo dissipation to $\ket{0}_\rm{imp}\ket{0}_1$, which is nonmagnetic.
Indeed, it is known that two-body losses remove pairs of fermions in a singlet state \cite{Nakagawa2020,Rey_Hot_reactive_fermions}---since only the singlet component of $\ket{\uparrow}_\rm{imp}\ket{\downarrow}_1$ is connected to $\ket{\uparrow\downarrow}_\rm{imp}\ket{0}_1$---thereby tending to suppress AF ordering.
Therefore, AF correlations are weaker with a higher dissipation rate $\gamma$.

In summary, our approach captures different occupation numbers and spin correlations describing the impurity cloud. Quantitatively, there is good agreement between our results and the ED results, clearly demonstrating the accuracy of our SGS ansatz. Furthermore, the approach is capable of showing the full crossover from small to large $\gamma$.

\subsection{Spin relaxation: crossover from the mean-field regime to the Kondo regime}

\begin{figure}[!t]
    \centering
    \includegraphics[width=0.8\linewidth]{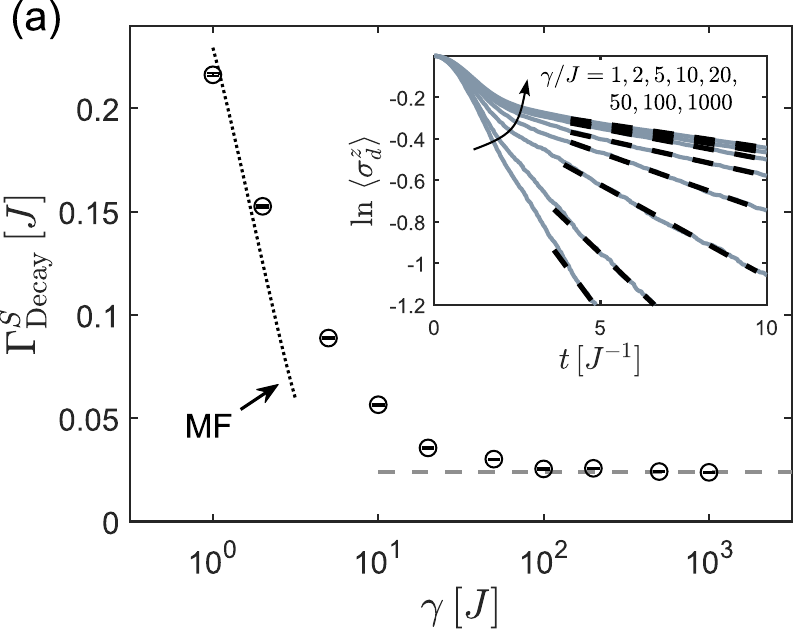}\\    
    \includegraphics[width=0.8\linewidth]{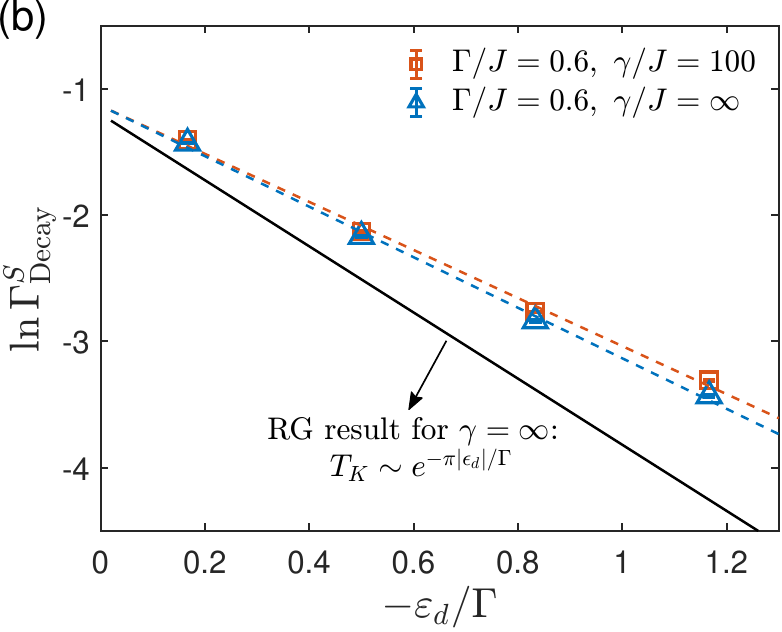}
    
    \caption{
    Spin relaxation. (a): the decay rate $\Gamma^S_\rm{Decay}$ (extracted from dynamics in the inset) versus dissipation rate $\gamma$. In the large $\gamma$ regime, the relaxation rate $\Gamma^S_\rm{Decay}$ converges, displaying a plateau as depicted by the dashed curve. This plateau indicates the Kondo regime with two-body losses. In the small $\gamma$ regime, the SGS results closely align with the mean-field results (dotted line). The inset shows the real time dynamics of  the impurity magnetization $\langle \hat{\sigma}_{d}^z\rangle=\langle \hat{n}_{d\uparrow}\rangle - \langle \hat{n}_{d\downarrow}\rangle$, which exhibits decay behavior $\langle \hat{\sigma}_{d}^z\rangle\sim e^{-\Gamma^S_\rm{Decay}t}$ at large times. The system size for the
 SGS simulations is set to $\ell=60$, with $\ce_d=-0.5J$ and $\Gamma=0.4J$.  (b): spin decay rate versus $\ce_d$.
The SGS results show that $\Gamma^S_\rm{Decay}$ exhibits an exponential dependence on $\ce_d/\Gamma$, $\Gamma^S_\rm{Decay}\sim \exp (-\eta \pi|\ce_d|/\Gamma)$, similar to the scaling of the Kondo temperature. A fitted coefficient of $\eta\sim0.61$ is obtained, while the RG prediction (solid curve) at $\gamma=\infty$ gives $\eta_\rm{RG}=1$.
 The discrepancy results from the renormalization of the effective spin interaction between the impurity site fermion and itinerant fermions by higher-order processes in $\Gamma/\ce_d$.}
    \label{fig:spin-relaxation}
\end{figure}

We now focus on the first Kondo signature: impurity’s spin relaxation slows down with increasing $\gamma$ \cite{short_paper}; in the Kondo regime, the lifetime of the impurity's magnetization is $1/T_K$, where $T_K$ is the Kondo temperature.
To capture spin relaxation with the SGS approach, we consider a larger system ($\ell=60$) with an onsite potential $\ce_d<0$. The chemical potentials are fixed to $\mu_L=\mu_R=0$, so the system is around half filling and only the symmetric modes need to be considered. To simulate the dynamics, we first turn off both dissipation and tunneling $\hat{H}_\rm{tun}$, i.e., set $\gamma=V=0$. The system is prepared as a disentangled static state $\ket{\uparrow}\ket{\rm{FS}}$, combining a spin-up impurity and the Fermi sea of the bath. Then we suddenly turn on both $\gamma$ and $V$, and the magnetization of the impurity, denoted by $\langle \hat{\sigma}^z_d\rangle = \langle \hat{n}_{d\uparrow}\rangle - \langle \hat{n}_{d\downarrow}\rangle$, starts relaxing toward the equilibrium value of zero.

In the inset panel of Fig.~\ref{fig:spin-relaxation}(a), we show the evolution of the logarithm of $\langle \hat{\sigma}^z_d\rangle$ with different dissipation rates $\gamma$. The dynamics exhibits two stages: the first stage is a universal decrease with the total tunneling rate $\Gamma\equiv\Gamma(\omega=0)=4V^2/J$, which is mainly due to the single-particle tunneling of charge to the bath. The second stage for $t>4 J^{-1}$ is a linear curve, i.e., an exponential decay of $\langle \hat{\sigma}^z_d\rangle$, with a rate dependent on $\gamma$,  which is fitted by using $\ln \langle \hat{\sigma}^z_d\rangle\sim -\Gamma^S_\rm{Decay}t + \mathcal{C}$ with the fitting parameter $\Gamma^S_\rm{Decay}$ representing the $\gamma$-dependent relaxation rate.

We emphasize here that we take the 
$\gamma=\infty$ limit by considering the quantum Zeno effect, which excludes double occupancy of the impurity and leads to unitary evolution without quantum trajectories.
As a result, the system in this limit is equivalent to that of the AIM with $U_d=\infty$, where the late-time spin relaxation is governed by the Kondo temperature $T_K\sim \sqrt{|\ce_d| \Gamma}\exp (- \pi|\ce_d|/\Gamma)$ \cite{Coleman}. 
As shown in the main panel of Fig.~\ref{fig:spin-relaxation}(a), the dynamics of the impurity magnetization slows down, in other words, $\Gamma^S_\rm{Decay}$ decreases with increasing $\gamma$.
Moreover, in the high dissipation regime ($\gamma>10^2J$), $\Gamma^S_\rm{Decay}$ slowly converges to a small but finite value, which we will demonstrate to be the Kondo temperature. Therefore, the convergence of $\Gamma^S_\rm{Decay}$ is a clear signature of the Kondo regime. In the small dissipation regime ($\gamma \lesssim 2J$), dynamics can be treated in a mean-field (MF) approximation (cf. Appendix \ref{appendix: HF}), as depicted by the dotted line.
The entire $\Gamma^S_\rm{Decay}$ curve represents a smooth crossover from the MF regime to the Kondo regime.

We now demonstrate that for extremely large $\gamma$, $\Gamma^S_\rm{Decay}$  exhibits an exponential dependence on $\ce_d/\Gamma$, similar to the scaling of the Kondo temperature. In Fig.~\ref{fig:spin-relaxation}(b), we plot $\ln \Gamma^S_\rm{Decay}$ as a function of $\ce_d/\Gamma$, showing SGS results for $\Gamma/J=0.6$ at $\gamma/J=100$ and $\gamma=\infty$. The data follows the form $\Gamma^S_\rm{Decay}\sim \exp (-\eta \pi|\ce_d|/\Gamma)$, with a fitting coefficient $\eta\sim0.61$.
In comparison, the renormalization group (RG) approach gives the same qualitative exponential form, but with a coefficient $\eta_\rm{RG}=1$ \cite{Coleman}. The discrepancy arises from the renormalization of the
effective spin interaction between the impurity site fermions
and itinerant fermions by higher-order processes in $\Gamma /\ce_d$.

In summary, our approach captures the full crossover of the spin relaxation from the MF regime to the Kondo regime. In the deep Kondo regime, our approach correctly exhibits the exponential form of the spin relaxation rate $\Gamma^S_\rm{Decay}$ (or the Kondo temperature $T_K$).

\begin{figure*}[ht] 
    \centering
    \includegraphics[width=0.92\linewidth]{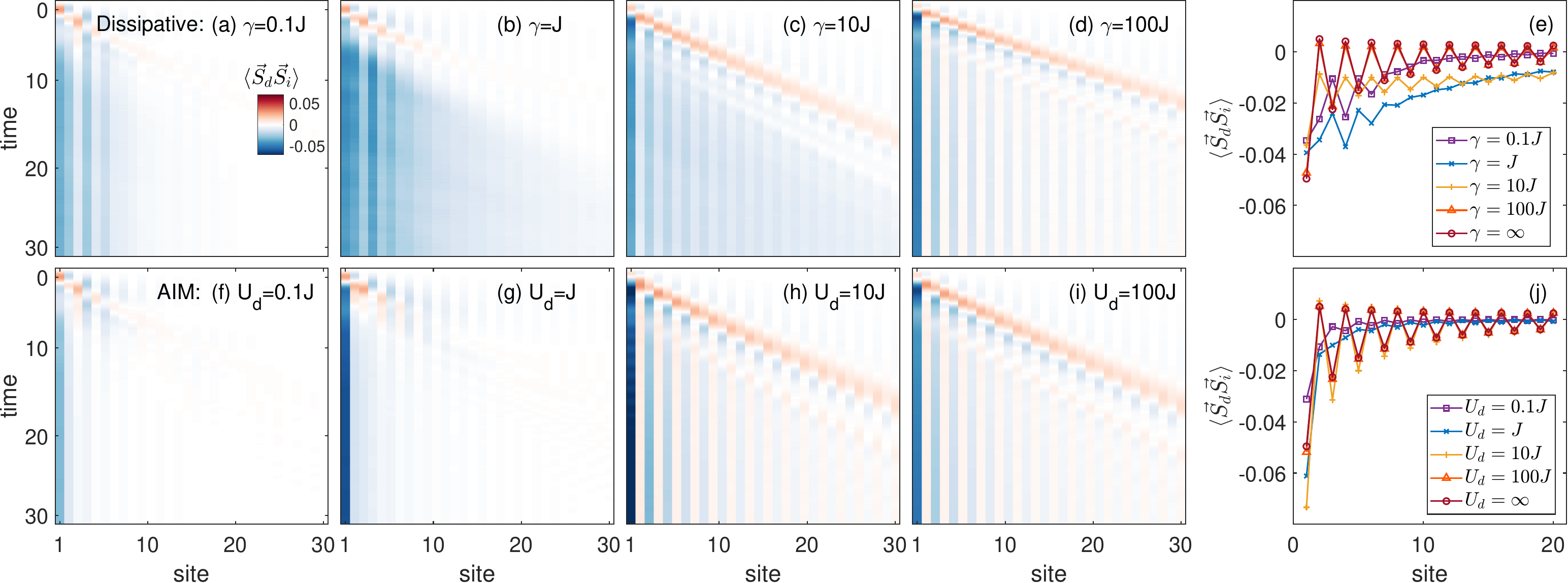}
    \caption{Dynamic Kondo cloud of the dissipative impurity model (a-d) and non-dissipative AIM (f-i). The system is initially prepared in a product state $\ket{\uparrow}\ket{\rm{FS}}$ with a system size of $\ell=60$, after which a dissipation rate $\gamma$ or repulsion coupling $U_d$ is applied at the impurity site.
    The transient spin correlations between the impurity site and the bath site $i$ (considering only symmetric modes) are computed using the SGS approach, with the time unit being $J^{-1}$.
    Panels (e) and (i) show the quasi-stationary spin correlations at time $t=20 J^{-1}$. Other parameters used are $\ce_d=-J, \Gamma=0.4J$.
    }
    \label{fig: transient spin cloud}
\end{figure*}

\subsection{ Transient spin correlations}

The previous analysis of the spin decay rate has focused primarily on properties of the impurity. However, the SGS approach also provides access to the correlations between the impurity and the bath. In this section, we will analyze the dynamics and quasi-stationary properties of the spin correlation functions, namely the properties of the so-called ``Kondo cloud”.

For comparison, we have performed simulations for both the dissipative impurity model \eqref{eq: model} and the non-dissipative AIM \eqref{eq: AIM}.  The onsite potential of the impurity is fixed to $\ce_d=-J$, and only the symmetric modes are considered. For the dissipative case, in Fig.~\ref{fig: transient spin cloud}(a-d) we show the spread of impurity-bath spin correlations $\langle \vec{S}_d \vec{S}_{i}\rangle$ starting from an initial state $\ket{\uparrow}\ket{\rm{FS}}$.
In accordance with physical intuition and with the Lieb-Robinson bound \cite{lieb1972} for dissipative systems, we observe the presence of a light-cone effect, i.e., most of the correlations emerge after the passing of a wavefront propagating at a constant speed.
The correlations at the wavefront are mostly ferromagnetic (FM), which can be understood as the initial magnetic moment of the impurity being carried away into the bulk of the fermionic reservoirs. Behind the first FM wavefront, correlations become mostly antiferromagnetic (AFM). We can clearly distinguish a low-dissipation ($\gamma/J=0.1,\,1$) from a high-dissipation regime ($\gamma /J=10,\,100$). In the former, the FM correlations at the wavefront decay rapidly in space and time, while a localized (non-propagating) region of AFM correlations is quickly stabilized around the impurity site. In contrast, at higher dissipation levels, the ferromagnetic correlations at the wavefront persist for an extended period, while the AFM spin correlations between the impurity and the bath gradually develop at the odd sites. The spin correlations at the even sites remain weak, with their signs shifting from negative to positive, i.e., FM. The spatially oscillating sign of impurity-bath correlations is a known property of equilibrium states of the Kondo model and AIM \cite{Nuss2015, Ashida2018a,Ashida2018b}.

We compare the above results with the non-dissipative case of the AIM, in which double occupancies at the impurity site are suppressed by a local repulsion $U_d$. The result of such computation is shown in Fig.~\ref{fig: transient spin cloud}(f-i). We can still distinguish a weakly interacting from a strongly interacting regime, with a FM light-cone. In the former, for $U_d/J=0.1,\,1$, we also observe a FM wavefront and an AFM zone around the impurity, but compared to the dissipative case, the wavefront decays quicker and the AFM correlations are tightly localized around the impurity site---essentially concentrated on the first bath site. For large interaction strengths $U_d/J=10,\,100$, the FM wavefront is well-defined, and the correlations within the light-cone display Kondo characteristics---stronger antiferromagnetic correlations at even sites and weaker ferromagnetic correlations at odd sites. We notice that in the dissipative case, this alternating sign behavior is established only at a very strong dissipation $\gamma/J=100$.

Finally, in Figs.~\ref{fig: transient spin cloud}(e) and \ref{fig: transient spin cloud}(j) we compare the near-stationary properties of the impurity-bath spin correlations, evaluated at a late time $t=20J^{-1}$. In general, we notice that the convergence to the Kondo regime $\gamma\to \infty$ or $U_d\to\infty$ (in which the two models coincide) is slower in the dissipative case than in the unitary one. When $\gamma/J=U_d/J=1$, the correlations are AFM on all sites, but the dissipative model exhibits stronger correlations compared to the AIM.
When $\gamma/J=U_d/J=10$, while the correlations in the AIM have already acquired the alternating sign typical of the Kondo regime, in the dissipative model they are still purely AFM. Moreover, the dissipative model shows lower AFM correlations at the first site than the dissipative model. We can interpret this observation as a consequence of the fact that the two-body losses remove fermions in a spin-singlet state, so that the dissipation tends to suppress the AFM correlations (as discussed in the main text).
In the Kondo regime ($\gamma/J=U_d/J=100$) the correlation functions are quite close to each other, with both cases displaying Kondo-type spin correlations with alternating signs between even and odd sites.

Although the correlations in the dissipative model eventually reproduce those of the AIM when the loss rate and the onsite repulsion become sufficiently large, the behavior at finite dissipation is markedly different from the Hamiltonian scenario at finite $U_d$.

\subsection{Charge conductance}

Another signature of the Kondo regime is the enhancement in charge conductance as the system progressively enters the Kondo regime. In our setup, we measure the transport through the impurity between biased reservoirs. In the absence of dissipation, transport is suppressed by an off-resonant $\ce_d<0$, while the Kondo effect increases the conductance with a high two-body loss rate.
Furthermore, transient currents and conductance can be accurately measured on recent experimental platforms, such as state-of-the-art cold-atom systems \cite{fabritius2024,mohan2024}. Therefore, we employ next the SGS approach to investigate the charge transport of the system transitioning from the MF regime to the Kondo regime.

Considering the lattice model \eqref{eq: lattice model} (where we set the charge unit to $e=1$), the time-dependent current from the reservoir to the impurity site is given by
\begin{equation}
    I_\alpha(t) = -\frac{dN_{\alpha}}{dt} = i V \sum_\sigma \rm{Tr} [( \hat{d}_{\sigma}^\dagger \hat{c}_{1 \sigma \alpha} - \hat{c}_{1 \sigma \alpha}^\dagger \hat{d}_{ \sigma})\hat{\rho}(t)],
    \label{current_definition}
\end{equation}
where $N_{\alpha}=\sum_{i\sigma} \rm{Tr} [\hat{c}^\dagger_{i\sigma\alpha} \hat{c}_{i\sigma\alpha}\hat{\rho}(t)]$ is the particle number of the left or right reservoir with $\alpha=L,R$. In the Kondo regime, a disentangled impurity takes a time of the order of $t_K\sim 1/T_K $ to form static antiferromagnetic correlations with the bath \cite{Nuss2015}.
The real-time simulations of the dynamics  are quite time-consuming as $t_K\sim\exp ( \pi|\ce_d|/\Gamma)$ grows exponentially with increasing $|\ce_d|$.
To obtain the equilibrium state efficiently, we first turn on an onsite repulsion $U_d=\gamma$ as in the AIM~\eqref{eq: AIM} and turn off the dissipation. By fixing $\mu_L=\mu_R=0$, the system is then prepared in the ground state (using the imaginary time evolution in simulations \cite{nondissipativeSGS}) of the AIM at half filling, in which a preliminary Kondo cloud is already formed.
We then turn off the onsite repulsion, turn on the dissipation, and quench the chemical potentials of the bath to $\mu_{L,R}=\mp \Delta\mathcal{V}/2$ with a potential bias $\Delta\mathcal{V}$ (which plays the role of a voltage bias in usual solid-state, quantum-dot setups), while keeping the number of particles in each reservoir fixed. The combination of losses and transport generates currents, and we measure the net flow of particles from the right to the left reservoir with $I(t,\Delta\mathcal{V})=(I_R-I_L)/2$. Note that this current does not distinguish between transported and dissipated particles. In the high dissipation regime, the transient charge conductance at zero potential bias, denoted by $G(t)=\frac{\partial I}{\partial\Delta\mathcal{V}}\big|_{\Delta \mathcal{V}=0}$, is anticipated to be proportional to the Kondo peak (of the impurity's spectral function) upon reaching equilibrium \cite{MeirWingreenFormula,transport_aim}. Consequently, the conductance is strongly connected to Kondo physics and provides a signature for the Kondo regime when the impurity potential is negative and off-resonant.

\begin{figure}[t]
  \centering  
  \includegraphics[width=0.8\linewidth]{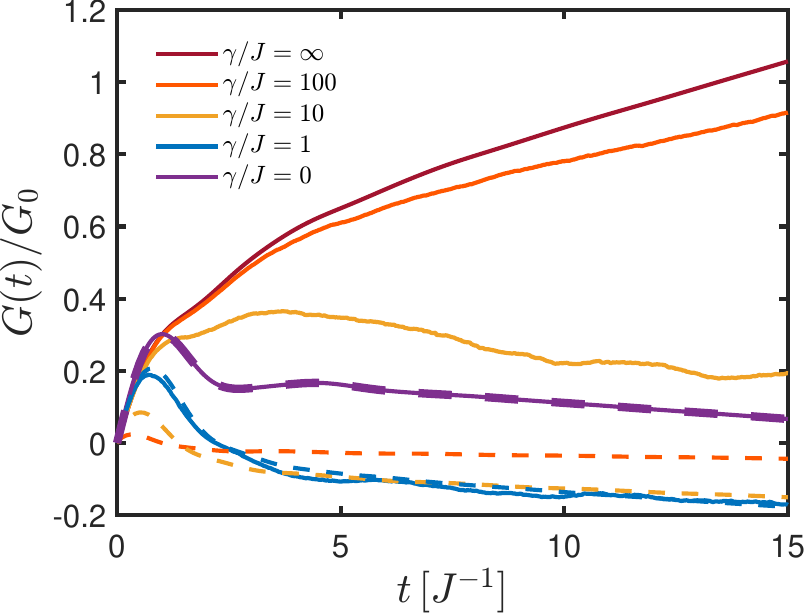}  
  
  \caption{Transient conductance for various dissipation rates $\gamma$. The signature of Kondo is an increase in charge conductance for transport through the impurity.
  We compare the SGS (solid curves) and MF (dashed curves) results, where $G_0=2/h$ represents the conductance quantum. In the low dissipation rate regime $\gamma\lesssim J$, the MF results agrees well with the SGS results, while the MF method fails in the high dissipation rate regime $\gamma\gtrsim 10J$. The parameters used are $\ell=40, \ce_d=-J, V=0.5J$ and $\Delta\mathcal{V}=0.001$. 
  }
  \label{fig: transport}
\end{figure}

In the following, we further fix the impurity potential to $\ce_d=-J$ and set the tunneling amplitude $V=0.5J$. Fig.~\ref{fig: transport} presents the transient conductance with zero potential bias and different loss rates $\gamma$.
For $\gamma=\infty$, the system becomes effectively closed due to the quantum Zeno effect, by which double occupations are excluded at the impurity site. The corresponding dynamics is equivalent to that of the AIM with $U_d=\infty$, which is in the Kondo regime with $\ce_d<0$.
During the evolution, the conductance increases and reaches the value $G_0=2/h$ (after restoring the Planck constant $h$) at $t \simeq 15 J^{-1}$.
A conductance of the order of the ``conductance quantum" $G_0$ is a signature of the Kondo regime, as it results from the emergence of the Kondo resonance near the Fermi surface \cite{Hewson}. At late times, we notice that the conductance tends to exceed $2/h$, which appears as a by-product of finite-size effects combined with variational errors from the SGS method \footnote{For a particle-hole symmetric AIM, where $\ce_d=-U_d$ in Hamiltonian \eqref{eq: AIM}, the SGS approach yields stable, quantized conductance of $2/h$ after several tunneling time intervals ($\Gamma^{-1}$). However, in the case of an asymmetric AIM, the conductance provided by the SGS approach continues to increase slightly beyond $2/h$, even for larger system sizes and extended simulation time.}.
For $\gamma=100J$, the conductance curve becomes slightly lower than that of $\gamma=\infty$, indicating that residual dissipation in the strong dissipation regime leads to a small suppression of the Kondo resonance \cite{short_paper}.
 When the loss rate further decreases to $\gamma=10J$, the conductance becomes highly suppressed with a value of $G \lesssim  G_0/4$ at large times ($t \gtrsim 10J^{-1}$).
Remarkably, the conductance reaches negative values when the loss rate $\gamma=J\sim\Gamma$ is intermediate.  We postpone the detailed discussion of this phenomenon to Sec.~\ref{sec: negative conductance}. 
When the loss rate decreases to $\gamma=0$, the conductance becomes positive again with a much lower value than $G_0$ because the impurity potential $\ce_d=-J$ is far off-resonant.

We also compare the MF results (shown as dashed lines in Fig.~\ref{fig: transport}) with the SGS results.
Without dissipation, both the MF and SGS approaches give the exact result. 
For small dissipation rates $\gamma=J$, the MF result agrees well with the SGS result, indicating that the dynamics are approximated by one-body losses and that the spin-up and spin-down sectors of the system are nearly decoupled. For large dissipation rates $\gamma\gtrsim 10J$, the MF approximations drastically deviate from the SGS results and fail to show the increased conductance with $\gamma=100J$ and $\gamma\to\infty$. The reason is that the entanglement between the spin-up and spin-down sectors becomes strong due to the dissipation-induced Kondo physics. However, this strong entanglement cannot be captured by MF approximations, in which the spin-up and spin-down sectors remain decoupled.

In summary, the SGS approach captures the conductance properties across all dissipation regimes.
In the high dissipation regime, our approach shows conductance enhancement in the Kondo regime. In the intermediate dissipation regime, our approach reveals the exotic phenomenon of ``negative conductance".

\section{Origin of ``negative conductance" \label{sec: negative conductance}}
In this section, we analyze the ``negative conductance" that appeared at intermediate dissipation rates $\gamma=J\sim\Gamma$.
Since the ``negative conductance" appears in the intermediate regime $\gamma\sim \Gamma$, dissipation can be effectively described by a mean-field approach in which the density matrix is assumed to be close to a Gaussian state. This ansatz approximates the original two-body loss with a self-consistent single-body loss (see Appendix \ref{appendix: HF}). The spin-dependent dissipation rates are given by $\gamma_\sigma=\gamma \langle \hat{n}_{d\bar{\sigma}}\rangle$, where $\bar{\sigma}$ denotes the opposite spin of $\sigma$, and $\langle \hat{n}_{d\bar{\sigma}}\rangle$ must be determined self-consistently. In the stationary state at late times, the self-consistency condition is
\begin{equation}
   \langle \hat{n}_{d\sigma} \rangle=-\ii\int\frac{\dd{\omega}}{2\pi} G^<_\sigma(\omega),
\end{equation}
where 
\begin{subequations} \label{eq: Green functions}
\begin{align}
G^{<}_\sigma (\omega) &=i \sum_\alpha f_\alpha(\omega) \Gamma_\alpha(\omega) \abs{G_\sigma^R(\omega)}^2,\\
    G_\sigma^R(\omega)&=\frac{1}{\omega-\ce_d-\Sigma_L(\omega)-\Sigma_R(\omega)+\frac{i}{2} \gamma_\sigma} 
\end{align}
\end{subequations}
are the lesser and retarded Green functions of the impurity.
Here, the self-energy for each channel is given by $\Sigma_{L,R}(\omega)=\Sigma(\omega\pm \Delta \mathcal{V}/2)$, with 
\begin{equation}
    \begin{aligned}
        \Sigma(\omega)&=\sum_p \frac{V_p^2}{\omega-\varepsilon_p+i0^+}\\
        &=\frac{V^2}{2J}\left(\frac{\omega}{J}-\sqrt{\frac{(\omega+i0^+)^2}{J^2}-4}\right)~.
    \end{aligned}
\end{equation}
The latter has been calculated using the expressions \eqref{eq: leads properties} (with $\mu_\alpha=0$), and quantifies the effect of the hybridization with the bath \cite{HaugJauho}.
Focusing on zero temperature, the Fermi-Dirac distribution reads $f_\alpha(\ce)=\vartheta(\mp\Delta\mathcal{V}/2-\ce)$ with the Heaviside step function $\vartheta(\ce)$.
The tunneling rates are determined through $\Gamma_{L,R}=-2\rm{Im}\Sigma_{L,R}$, whose explicit expressions are given in Eq.~\eqref{eq: tunneling rates}.
Through the Keldysh formalism, the static current of the $\alpha$-lead is given by the Meir-Wingreen formula \cite{MeirWingreenFormula}:
\begin{equation} \label{eq: channel current}
    I_\alpha =\frac{i}{h}\sum_\sigma\int d \ce\,\Gamma_\alpha(\ce) \left[ 2 i f_\alpha(\ce) \rm{Im} G^R_\sigma(\ce) + G^{<}_\sigma(\ce)\right].
\end{equation}
Using Eqs.~\eqref{eq: Green functions} and \eqref{eq: channel current}, the imbalanced current $I\equiv(I_R-I_L)/2= I_\rm{ch}+I_\rm{loss}$ can be regarded as having two contributions \cite{gievers2023}. The first one
\begin{equation}
    I_\rm{ch} = \frac{1}{h}\int d \epsilon \left(f_R-f_{L}\right) \Gamma_L \Gamma_{R} \sum_\sigma \abs{G_\sigma^R}^2
\end{equation}
receives contributions only at the edges of the band of the baths since the pre-factor $f_R-f_{L}$ is nonzero only at the edges.
The second current
\begin{equation} \label{eq: loss currrunt}
    I_\rm{loss} =  \frac{1}{2h}\int d \epsilon \,(f_R\Gamma_R-f_L\Gamma_L) \sum_\sigma \gamma_\sigma\abs{G_\sigma^R}^2
\end{equation}
mostly results from the effective one-body losses of the impurity.
Correspondingly, the conductance $G =  G_\rm{ch}+G_\rm{loss}$ at zero potential bias is divided into two parts with $G_X=dI_X/d\Delta\mathcal{V}, X=\rm{ch},\rm{loss}$.
\begin{figure}[t]
  \centering  
  \includegraphics[width=0.85\linewidth]{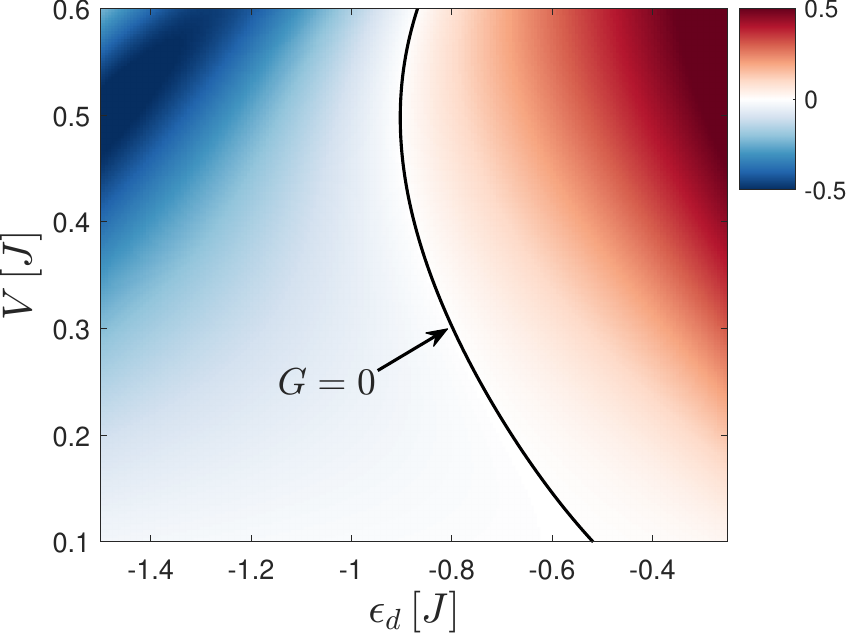}
  \caption{Stationary conductance with the MF approximation at intermediate dissipation $\gamma=J$. The units of energy and conductance are $J$ and $G_0=2/h$, respectively.}
  \label{fig: reverse current through GS appr.}
\end{figure}
We find the main contribution to the negative $G$ is the loss part $G_\rm{loss}$. Fig.~\ref{fig: reverse current through GS appr.} shows the density plot of the total conductance $G$ as a function of $\ce_d$ and $V$. The borderline of $G=0$ (solid black curve) is near $\ce_d\sim -J$, whose value of $\ce_d$ slightly increases with increasing $V$.
The negative conductance appears in the left regime and becomes larger with increasing $V$ and $\abs{\ce_d}$.

Physical insight into the ``negative conductance" can be gained by considering a simpler case where the density of states $\Gamma_{L,R}(\ce)=\Gamma$ is constant within the band $\ce \in[-2J\mp \Delta\mathcal{V}/2,2J\mp \Delta\mathcal{V}/2]$. The analytic expression of the aforementioned $G_\rm{loss}$ then explicitly reads
\begin{align}
    G_\rm{loss}=\frac{1}{2h}\sum_\sigma \Bigg[&\frac{\gamma_\sigma}{\ce_d^2+(\Gamma+\gamma_\sigma)^2} \notag\\
    &- \frac{\gamma_\sigma}{(\ce_d+2J)^2+(\Gamma+\gamma_\sigma)^2}\Bigg],
\end{align}
which clearly depends on the difference of the two band edges in Eq.~\eqref{eq: loss currrunt}.
It is straightforward to show that $ G_\rm{loss}<0$ if $\ce_d<-J$. The physical reason is that the impurity potential $\ce_d<-J$ is closer to the lower edge of the band. Therefore, the left reservoir has a larger (loss) current as a result of the one-body dissipation of the impurity.

To summarize, such ``negative conductance" is a one-body phenomenon that manifests only in the presence of the aforementioned potential bias and a narrow-band scenario, in which the magnitude of the (negative) impurity potential $\abs{\ce_d}$ is comparable to (or even larger than) the bandwidth.
\vspace{1cm}

\section{Realization of multi-channel spin-1 Kondo physics} \label{sec: high spin}
In this section, we discuss a simple generalization of the dissipative setup discussed so far that allows the realization of a Kondo model with spin higher than $1/2$. This setup has already been briefly discussed in the companion paper \cite{short_paper}.
\subsection{Analytic discussion}
We consider a generalized model in which the “dot” is formed by two sites of the chain, subject to two-body losses [as depicted in Fig.~\ref{fig: FM domain}(a)]:
\begin{equation}
    \dt\hat{\rho}=-\ii\comm{\hat{H}}{\hat{\rho}}+\sum_{a=0,1}\frac{\gamma_a}{2}(2\hat{L}_a\hat{\rho} \hat{L}_a^\dag-\{\hat{L}_a^\dag \hat{L}_a,\hat{\rho}\}),
\end{equation}
where
\begin{subequations}\label{eq: model with two impurity sites}
    \begin{align} 
    & \hat{H}=\hat{H}_\rm{imp}+\hat{H}_\textup{leads}+\hat{H}_\textup{tun} ,\\    
    & \hat{H}_\rm{imp} = \sum_{\sigma} \big[\ce_d\sum_a \hat{d}_{a\sigma}^\dag \hat{d}_{a\sigma}  \notag\\    
    &\qquad\qquad\quad - V (\hat{d}_{0\sigma}^\dag \hat{d}_{1\sigma}+\rm{H.c.}) \big] ,\\    
    & \hat{H}_\rm{leads}=-\sum_{j\sigma\alpha} J_{\alpha} (\hat{c}_{j\sigma\alpha}^\dag \hat{c}_{j+1\sigma\alpha }+\rm{H.c.}),\notag\\   
    & \hat{H}_\rm{tun}=-\sum_{a\sigma\alpha}(\bar{J}_{a \alpha} \hat{d}_{a\sigma}^\dag \hat{c}_{1\sigma\alpha}+\rm{H.c.}) ,\\
    & \hat{L}_a =  \hat{d}_{a\uparrow} \hat{d}_{a\downarrow} \label{eq: jump 2} .
\end{align}
\end{subequations}
Here, different from the lattice model \eqref{eq: lattice model} (with one impurity site), $V$ and $J_\alpha$ denote the hopping amplitudes within the impurity sites and the bath sites, respectively. $\bar{J}_{a\alpha}$ represents the tunneling amplitudes between the impurity site $a$ and the reservoir $\alpha$. In addition, both impurity sites have two-body losses with dissipation rates $\gamma_a$, denoted by jump operators $\hat{L}_a$.
As discussed in \cite{short_paper}, the strategy that we follow is to tune the parameters such that the dark states \cite{BreuerPetruccione} of the set $\{\hat{H}_\textup{imp},\,\hat{L}_0,\,\hat{L}_1\}$---which would be stationary in the absence of the leads---form a low-energy spin-1 multiplet. In the case of \eq~\eqref{eq: model with two impurity sites}, the dark subspace consists of four sets of states: the empty impurity state $\ket{\mathcal{N}=0,\mathcal{S}=0, \mathcal{M}=0}$ with zero energy, two doublets $\ket{\pm,\mathcal{N}=1,\mathcal{S}=1/2, \mathcal{M}}$ with energy $\ce_d\mp V$ and a triplet $\ket{\mathcal{N}=2,\mathcal{S}=1, \mathcal{M}}$ with energy $2\ce_d$, where $\mathcal{N}$ and $\mathcal{S}$ denote the particle number and the total spin, respectively, and the magnetization takes the values $\mathcal{M}=-\mathcal{S},\dots, \mathcal{S}$. As noted in \cite{short_paper}, the bright states (i.e. the ones outside of the dark subspace) of the dissipative, two-site dot have nontrivial properties---in particular, the slowest decay rate $\gamma_\textup{min}$ of the bright states is non-monotonic in $\gamma$ (with fixed $V$), and goes to zero both for $\gamma\to0$ and $\gamma\to+\infty$. Therefore, to ensure that dissipation suppresses the population of all states except the dark subspace, one needs to work close to the maximal value of $\gamma_\textup{min}\propto\gamma$, which occurs for $\gamma\propto V$. Summarizing, if we have
\begin{subequations}
\begin{align}
& \gamma \sim V \sim |\ce_d|~,\label{condition: 1}\\
&\Delta \ce=-\ce_d - V > 0, \label{condition: 2}
\end{align}  
\end{subequations}
then for times larger than a few $\gamma^{-1}$ the dynamics will be confined to the dark subspace \footnote{We notice that if $\gamma\gg V$ but $V$ is sufficiently larger than the bandwidth of the leads, the transitions to the bright states are highly off-resonant and thus suppressed---but in a more usual, Hamiltonian way.}. The second condition \eqref{condition: 2} ensures that the lower spin-$1/2$ doublet has a higher energy than the spin-$1$ triplet. Since we are interested in the limit $V\sim\gamma\to+\infty$, we also need to require $\abs{\ce_d}\sim V$ to keep $\Delta\ce$ finite and allow particle exchanges with the leads.  In the spirit of adiabatic elimination \cite{Garcia-Ripoll_2009,Kessler}, to the lowest order in $\gamma^{-1}$, the effective dynamics of the system will be projected onto the dark subspace, even after turning on the tunneling $\bar{J}$ with the leads. This construction leads to an effective Hamiltonian dynamics governed by an ionic model \cite{Hewson,Bickers,NozieresBlandin}.
If the coupling to the leads is sufficiently weak,  
\begin{equation}
| \Delta \ce | \gg \Gamma \sim \frac{\bar{J}^2}{J},\label{condition: 3}
\end{equation}
the effective model can be mapped via a \emph{unitary} Schrieffer-Wolff transformation to a spin-$1$ Kondo model:
\begin{equation}\label{eq: spin 1 Kondo}
    \hat{H}_{S=1}=\sum _{\alpha\beta} \frac{\tilde{J}_{\alpha} \tilde{J}_{\beta }}{2 | \Delta \ce | }\vec{S}_{\rm{imp}}\vec{s}_{\alpha  \beta }+\hat{H}_\textup{leads}.
\end{equation}
In the expression above, we have introduced the $SU(2)$ impurity spin $\vec{S}_\textup{imp}$ belonging to the spin-$1$ representation and the bath spin-$1/2$ operators $\hat{s}_{\alpha  \beta }^{x,y,z}=\frac{1}{2}  \hat{c}_{1 \sigma\alpha}^{\dagger} \tau^{x,y,z} _{\sigma \sigma'} \hat{c}_{1 \sigma'\beta }$. The notation $\tilde{J}_{\alpha }=\sum_a(-1)^a f_a \bar{J}_{\alpha  a}$, with $f_0=1$, $f_1=\rm{sign}(V)=\pm 1$,  corresponds to the ground-state wave functions of $H_\rm{imp}$ in the one-particle sector. Note that if $V$ is positive and the hopping amplitudes $\bar{J}_{a\sigma}$ are the same, all $\tilde{J}_{\alpha}$ become zero due to the parity, and the effective Hamiltonian \eqref{eq: spin 1 Kondo} thus vanishes. Physically \cite{Hewson,Coleman}, \eq~\eqref{eq: spin 1 Kondo} describes a situation where the low-lying spin triplet state is always occupied (i.e. the impurity is always hosting two fermions) and behaves like a spin $1$ object, while the lead-mediated transitions to the spin-$1/2$ states are virtual, and cause an antiferromagnetic interaction between $\vec{S}_\textup{imp}$ and the lead fermions. We observe that, with the same tuning of the loss rate and hopping as above, but with $\Delta \ce < 0$, we can map the effective, strong-dissipation dynamics to a spin-$1/2$ Kondo model:
\begin{equation}\label{eq: spin half Kondo}
    \hat{H}_{S=\tfrac{1}{2}}=\sum _{\alpha \beta a b} \frac{\tilde{J}_{\alpha  a} \tilde{J}_{\beta  b}}{2 \sqrt{2} | \Delta \ce | }\vec{S}_{\rm{imp}}\vec{s}_{\alpha  \beta }+\hat{H}_\textup{leads}~,
\end{equation}
where $\vec{S}_\textup{imp}^2=3/4$.
\par It is easy to see that the coupling to the leads will generally endow the dark states of $H_\textup{imp}$ with a finite lifetime, an effect that is captured by the first-order terms in the dissipative Schrieffer-Wolff transformation, analogous with the scenario of a single dissipative site. While we will not pursue this calculation here, it is straightforward to estimate that the spin-$1$ states will suffer from a decay rate of order $\bar{J}^2/\gamma$, since they are coupled to the bright states by the tunneling of a single fermion from the leads. Drawing on the results of \cite{short_paper} for the single-site case, we can then expect that the competition between the emerging Kondo physics and the residual dissipation will allow observation of the former as long as  
\begin{equation}
    \frac{\bar{J}^2}{\gamma }\lesssim T_K\sim \sqrt{\Gamma  | \Delta \ce | }\ee^{- \pi\frac{ | \Delta \ce | }{\Gamma }},\label{condition: 4}
\end{equation}
where $T_K$ is the Kondo temperature. It is interesting to note that the spin-$1/2$ dark states are connected to the bright states by two tunneling events, hence they will have a suppressed decay rate $\bar{J}^4/\gamma$. Therefore, in the case of $\Delta\ce<0$, the effective Hamiltonian \eqref{eq: spin half Kondo} would be more protected from the residual dissipation.

\subsection{Stability of impurity ferromagnetism}
\begin{figure}[t]
    \centering    \includegraphics[width=\linewidth]{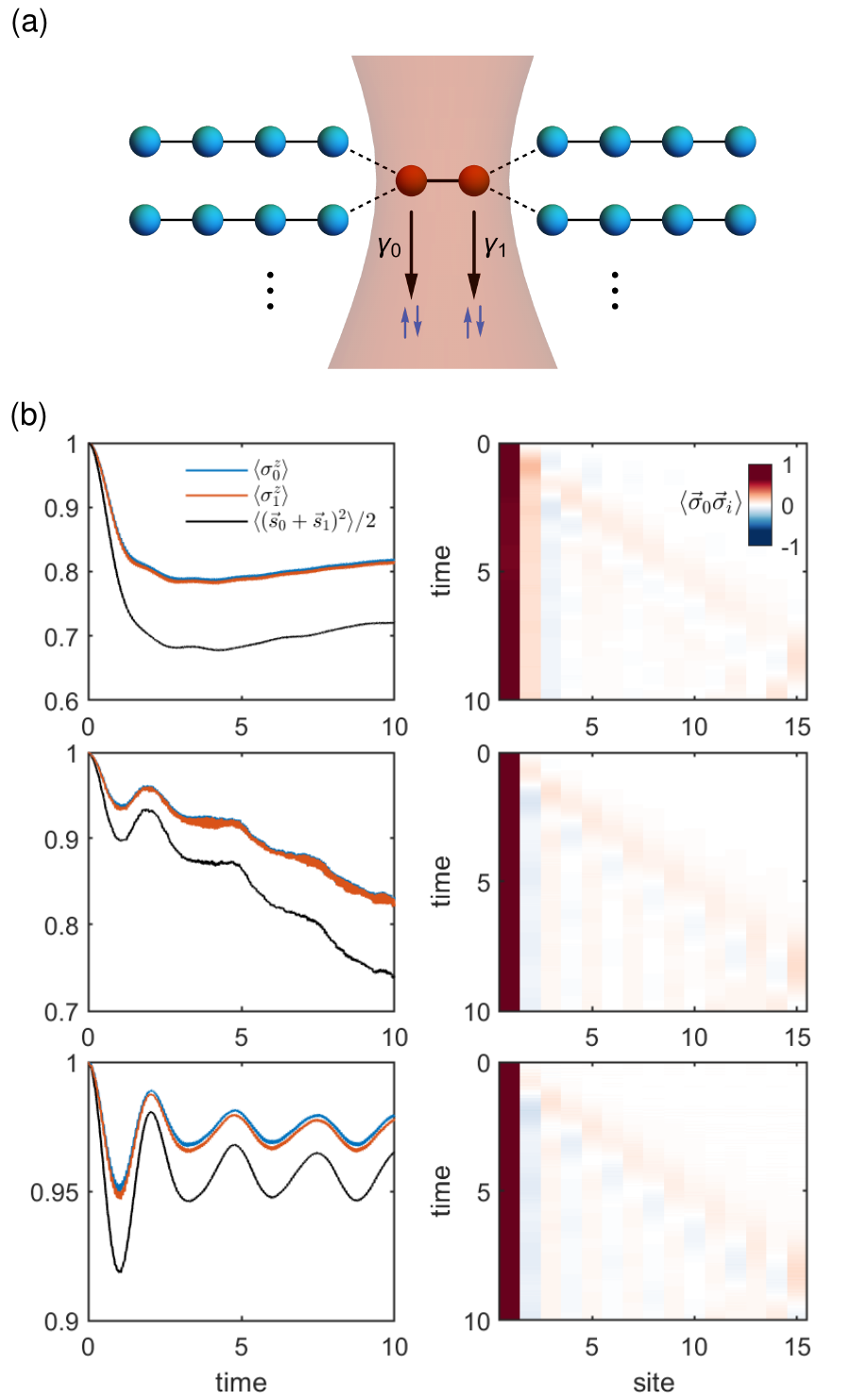}

    \caption{(a) Schematic depiction of the generalized model, with two impurity sites undergoing two-body losses.
    (b) Stabilization of a ferromagnetic impurity,  observed through SGS simulations. The right panel shows the transient spin correlations $\langle\vec{\sigma}_0\vec{\sigma}_i\rangle$ with $\gamma=0,\, V,\, \infty$ (from top to bottom) and the initial state $\ket{\uparrow\uparrow}_d \ket{\rm{FS}}_l$. The time unit is $J^{-1}$ and the parameters used are $V=50J,\Delta\ce=2J, \ell=15$.  The FM domain is stabilized by large dissipation.
    }
    \label{fig: FM domain}
\end{figure}

In the previous section, we delineated a possible mapping from a model with two dissipative sites to a spin-$1$ Kondo model. Essentially, this mapping relies on the ability of two-body losses to stabilize a ferromagnetic (FM) region on the impurity sites. The triplet of spin-$1$ states mentioned before represent precisely these kind of states. More generally, a FM behavior is generically expected from two-body losses even in longer dissipative chains \cite{Rey_Hot_reactive_fermions}. 

In this Section, we corroborate the existence of FM intra-dot correlations in the model \eqref{eq: model with two impurity sites} of two strongly coupled impurity sites, in which only one is subjected to two-body losses $\gamma_0=\gamma$, $\gamma_1=0$ while the other is connected to a single lead. To probe the stability of impurity ferromagnetism, we have run our SGS simulations with the initial state $\ket{\uparrow \uparrow}_d\ket{\rm{FS}}_l$, where $\ket{\uparrow \uparrow}_d=\hat{d}_{0\uparrow}^\dag \hat{d}_{1\uparrow}^\dag\ket{0}_d$ is an initial, spin-polarized impurity state and $\ket{\rm{FS}}_l$ is a half-filled Fermi sea of the lead (having $\ell=15$ sites) with zero net magnetization. In \fig~\ref{fig: FM domain}(b) we show the results of the SGS calculations with increasing dissipation $\gamma=0,\, V,\, \infty$ and for strongly coupled impurity sites $V=50J$. The onsite energies $\ce_d$ are chosen such that $\Delta\ce=2J$ is smaller than the bandwidth of the lead, thus allowing for resonant exchanges of particles.
The left panels of the Figure show the time evolution of the average spin orientations $\langle\hat{\sigma}_0^z\rangle$ and $\langle\hat{\sigma}_1^z\rangle$ at the two impurity sites, and the magnitude $\langle\vec{S}_\textup{imp}^2\rangle$ of the total spin, $\vec{S}_\textup{imp}=\vec{s}_0+\vec{s}_1$.
The overall magnitude of $\langle\hat{\sigma}_{0,1}^z\rangle$ decreases in time for both $\gamma=0$ and $\gamma=V$, albeit in quite different ways---most recognizably, the behavior in the former case is not monotonic. For $\gamma=\infty$, the dynamics of $\langle\hat{\sigma}_{0,1}^z\rangle$ is nearly “frozen”, namely both spins fluctuate only slightly with respect to their initial value. The behavior of the total impurity spin magnitude $\langle\vec{S}_\textup{imp}^2\rangle$ mirrors the local magnetizations: it decays significantly for $\gamma=0$ and $V$, while it oscillates around $1.9$ for $\gamma\to\infty$---quite close to the perfect spin-$1$ value $\langle\vec{S}_\textup{imp}^2\rangle=2$. Thus, we observe that the combination of a strong coupling $V$ between the impurity sites and dissipation enforces collective dynamics of the total impurity spin.

The right panels of \fig~\ref{fig: FM domain}(b) show the spin correlations $\langle\vec{\sigma}_0\vec{\sigma}_i\rangle$ between the dissipative site $i=0$ and the other sites, i.e. the second impurity site $i=1$ and the lead sites $i\ge2$. These plots confirm the establishment of a strong FM correlation between the two impurity sites, which is further stabilized by increasing $\gamma$. The correlations in the rest of the system are far weaker, and emerge after the passage of a slightly FM light-cone.  Notably, the correlations with the lead sites $i>1$ exhibit alternating signs for $\gamma>0$, starting with an AFM correlation for $i=2$, while for $\gamma=0$ there is a small ``leakage” of FM correlations to the first bath site.

Summing up, these preliminary simulations confirm the picture we presented in the previous section---that is, two impurity sites can be tuned to behave like a spin-$1$ impurity by suitably choosing their energies, tunneling rates, and dissipation. With the present choice of parameters, a very large loss rate $\gamma$ appears to be the most suitable setup for obtaining an effective spin-$1$, as it gives a value of $\langle\vec{S}_\textup{imp}^2\rangle$ closer to $2$. This conclusion might seem at odds with the the previous section (cf. also \cite{short_paper}), where we mentioned that for $\gamma\to+\infty$ the dark subspace includes also other non-FM states besides the Dicke multiplets. However, for the large value of $V$ chosen in this section, these states are extremely off-resonant with respect to the transitions with the bath and thus can be considered to be excluded from the dynamics.

\section{Conclusions and outlook} \label{sec: conclusions}

In this work, we introduced a non-perturbative framework that combines the SGS variational ansatz with the quantum trajectory approach, offering an efficient method for simulating dissipative impurity systems.
We applied this method to a system where a spinful impurity, subject to two-body losses, interacts with a bath of noninteracting fermions. Through this approach, we demonstrated that strong localized two-body losses can induce the Kondo effect, consistent with findings in related research \cite{short_paper}. The key signatures of Kondo physics, such as the slowdown of spin relaxation and the enhancement of charge conductance, were effectively captured using the SGS framework. This method allowed us to thoroughly examine the full crossover from the weakly correlated mean-field regime at small dissipation to the strongly correlated Kondo regime at large dissipation, thereby providing a comprehensive understanding of the system’s behavior across different dissipation regimes. Notably, in the mean-field regime, we revealed the emergence of exotic ``negative conductance" at zero potential bias, which arises from intermediate dissipation and the constraints of a finite reservoir bandwidth.

Our framework can be extended to more complex scenarios, including systems with multiple dissipative impurity sites, where interactions and losses can occur simultaneously at the same site, or involving a superconducting or Mott-like fermionic baths. Our exploration of dissipative Kondo models suggests new directions for utilizing localized dissipation to potentially realize large-scale, tunable spin-1/2 and higher-spin Kondo systems in cold-atom setups. These systems could serve as a basis for investigating strongly correlated phenomena within the broader Kondo problem family, such as quantum criticality \cite{Jones1988,Affleck1992,Affleck1993,bayat2014,keller2015,iftikhar2015,iftikhar2018,lorenzo2017quantum,pouse2023}, fractionalization \cite{Emery1992,Gan1995,Sela2011,Mitchell2012,Mitchell2012_2,Landau2018,Karki2023}, and multistage Kondo effects \cite{Hofstetter2001,Pustilnik2001,Wiel2002,Granger2005,Karki2018,guo2021}.

\section{Acknowledgments}
We acknowledge useful discussions with W. Zhang, D. Sels, A. Gomez Salvador, R. Andrei, M. Kiselev. Y.Q. and E.D. acknowledge support by the SNSF project 200021\_212899, NCCR SPIN, a National Centre of Competence in Research, funded by the Swiss National Science Foundation (grant number 225153), the Swiss State Secretariat for Education, Research and Innovation (contract number UeM019-1). E.D. also acknowledges support from the ARO grant number W911NF-20-1-0163. T.E., Y.Q., and E.D. acknowledge funding by the ETH grant. M.S. and J.M. have been supported by the DFG through the grant HADEQUAM-MA7003/3-1. J.M. acknowledges the Pauli Center for hospitality. The numerical simulations were performed on the ETH Euler cluster.

\appendix

\section{Calculation of $\hat{c}'\ket{\rm{GS}_s}$ and $\hat{c}'^\dagger\ket{\rm{GS}_s}$ \label{appendix: manipulation of GS}}
In this appendix, we provide analytical details on \eq~\eqref{eq: new GS}---namely, on how to calculate  $\hat{c}'\ket{\rm{GS}_s}$ and $\hat{c}'^\dagger\ket{\rm{GS}_s}$ given the Gaussian state $\ket{\rm{GS}_s}$ and $\hat{c}'^\dagger=\sum_{j=1}^{N_f} v_j \hat{C}_j^\dagger$. We recall that the Gaussian state is expressed as a single Slater determinant, given by $\ket{\rm{GS}_s}=\hat{C}_1'^\dagger \hat{C}_2'^\dagger \cdots \hat{C}_{N_s}'^\dagger \ket{0}$. Here, the operators $\hat{C}_i'^\dag=\sum_{j=1}^{N_f} u_{ij} \hat{C}_j^\dag$, where $i=1,\dots,N_s$ represent the occupied modes, and the $N_f\times N_s$ matrix $u$ contains the associated wavefunctions.

To calculate $\hat{c}'^\dagger\ket{\rm{GS}_s}$, we decompose the creation operator $\hat{c}'^\dag$ as follows:
\begin{equation} \label{eq: c bar dagger}
\hat{c}'^\dag = \beta\, \hat{c}''^\dag + \sum_{i=1}^{N_s} f_i \hat{C}_i'^\dag.
\end{equation}
Here, the first term represents a normalized mode $\hat{c} ''^\dag=\sum_{j=1}^{N_f} \tilde{v}_j \hat{C}_j^\dagger$, where $\tilde {\mathbf{v}}=(\mathbf{v}-u u^\dag \mathbf{v})/\beta$ is orthogonal to the occupied modes of $\ket{\rm{GS}_s}$, and is weighted by $\beta=\norm{\mathbf{v}-u u^\dag \mathbf{v}}$. In the previous expressions, $\bm{v}$ is the vector collecting the coefficients $v_j$. The second term is a linear combination of the occupied modes with coefficients $f_i=\sum_{j=1}^{N_f} u_{ij}^*v_j$, which annihilates $\ket{\rm{GS}_s}$ as a consequence of the Pauli exclusion principle. Therefore, $\hat{c}'^\dagger\ket{\rm{GS}_s}$ simplifies to:
\begin{equation}
    \hat{c}'^\dagger\ket{\rm{GS}_s}=\beta\, \hat{c}''^\dag \hat{C}_1'^\dagger \hat{C}_2'^\dagger \cdots \hat{C}_{N_s}'^\dagger \ket{0}\equiv \beta \ket*{\small\widetilde{\rm{GS}}_s}.
\end{equation}
The last equality defines a new Slater determinant and hence a new Gaussian state. 

Next, to calculate $\hat{c}'\ket{\rm{GS}_s}$, we notice that only the second term in Eq.~\eqref{eq: c bar dagger} contributes, as it represents a superposition of the occupied modes. We recombine the occupied modes into $\hat{C}_i''^\dag=\sum_{i'=1}^{N_s}\omega_{ii'} \hat{C}_{i'}'^\dag, i=1,\dots,N_s$. where $\omega$ is a unitary $N_s\times N_s$ matrix whose first column is $\mathbf{f}/\norm{\mathbf{f}}$ ($\mathbf{f}$ is the vector of the coefficients $f_j$ defined in the previous paragraph) and the remaining columns are orthogonal to $\bf f$. The ambiguity in the definition of $\omega$ is harmless, since it will correspond to the choice of the overall phase of $\hat{c}'\ket{\rm{GS}_s}$. In terms of the new modes $\hat{C}_i''$, the Gaussian state can be expressed as $\ket{\rm{GS}_s}=\rm{det}(\omega)^{-1} \hat{C}_1''^\dagger \hat{C}_2''^\dagger \cdots \hat{C}_{N_s}''^\dagger \ket{0}$ and $\hat{c}'^\dag=\beta \hat{c}''^\dag+\norm{\mathbf{f}}\hat{C}_1''^\dag$, from which we see that the operator $\bar{c}$ simply removes the first mode $\hat{C}_1''^\dagger$ from the state $\ket{\rm{GS}_s}$ . Therefore, we obtain
\begin{equation} \label{eq: c GS}
    \hat{c}'\ket{\rm{GS}_s}=\beta'\, \hat{C}_2''^\dagger \cdots \hat{C}_{N_s}''^\dagger \ket{0}\equiv \beta' \ket*{\small\widetilde{\rm{GS}'}_s},
\end{equation}
which defines a new Gaussian state, with the coefficient $\beta'=\norm{\mathbf{f}}/\rm{det}(\omega)$.

\section{Evolution of the state after a quantum jump\label{appendix: evolution after jump}}

In this appendix, we describe the procedure for determining the evolution of the quantum state immediately following a quantum jump. The simple projection rules \eqref{eq: quantum jump} leave the bath's Gaussian states corresponding to the singly and doubly occupied impurity undefined. We will show how to resolve this ambiguity by performing a short-time evolution of the post-jump state under the non-Hermitian Hamiltonian $H_{\rm{NH}}$.

Assume a quantum jump occurs at time $t_1$ in a quantum trajectory. Immediately after the jump, the state of the trajectory  is given by $\ket{\Psi(t_1^+)} = \ket{0}\ket{\rm{GS}_0}$, where the Gaussian state $\ket{\rm{GS}_0}$ can be represented by the unitary matrix $U_0(t_1^+)=U_2(t_1^-)$, as described in Sec. \ref{sec: SGS techniques}. Our task is to obtain the new matrices $U_{\uparrow,\downarrow,2}(t_1^+)$ after the jump. We evolve the state $\ket{\Psi(t_1^+)}$ for a small time interval $\delta t$ under the non-Hermitian Hamiltonian $H_{\rm{NH}}$, using a Taylor expansion:
\begin{equation} \label{eq: Taylor expansion}
\begin{aligned}
        \ket{\Psi(t_1+\delta t)} &= \exp(-i \hat{H}_\rm{NH}\delta t )\ket{\Psi(t_1^+)}\\
        &=[1-i \hat{H}_\rm{NH}\delta t-\frac{1}{2} \hat{H}^2_\rm{NH}\delta t^2 + o(\delta t^2)]\ket{0}\ket{\rm{GS}_0} \\
        &\approx \ket{0}\ket{\rm{GS}_0}+ \sum_{s'=\uparrow,\downarrow,2} \alpha_{s'} \ket{s'}\ket{\rm{GS}_{s'}}.
\end{aligned}
\end{equation}
Since the action of $\hat{H}_{\rm{NH}}$ can transfer at most one fermion from the reservoirs to the impurity, we need to keep terms up to the second order to recover all four impurity states. By retaining the leading term for each impurity subsector, we can represent $\ket{\Psi(t_1+\delta t)}$ as an SGS in the third line of Eq.~\eqref{eq: Taylor expansion}. The Gaussian states and their coefficients are given by
\begin{equation} 
\begin{aligned}    \alpha_\uparrow \ket{\rm{GS_\uparrow}}&=-i\delta t \,\hat{c}'_\uparrow\ket{\rm{GS}_0},\\
\alpha_\downarrow\ket{\rm{GS_\downarrow}}&=i\delta t \,\hat{c}'_\downarrow\ket{\rm{GS}_0},\\
\alpha_2\ket{\rm{GS_2}}&=\delta t^2\,\hat{c}'_\uparrow\hat{c}'_\downarrow\ket{\rm{GS}_0},
\end{aligned}
\end{equation}
which can be calculated using Eq.~\eqref{eq: c GS}. The first $N_{s'}$ (representing the number of bath particles in the impurity subsector $s'$) columns of the corresponding matrix $U_{s'}$ are determined by the wavefunctions of the occupied modes in $\ket{\rm{GS}_{s'}}$. The remaining columns of $U_{s'}$ are chosen to be orthonormal to the first $N_{s'}$ columns. The updated SGS, as described in Eq.~\eqref{eq: Taylor expansion}, is then used to continue the simulation of real-time evolutions through the SGS approach, utilizing the variational parameters $\alpha_s$ and $U_s$.
\vspace{0.5cm}

\section{Gaussian state approach}\label{appendix: HF}
In this appendix, we present the Gaussian state (GS) approach \cite{shi2018}, a self-consistent mean-field method, applied to the dissipative impurity model described by Eq.~\eqref{eq: model}.

As stated in the main text, we approximate the density matrix of the system with a Gaussian density matrix $\rho_{\rm{GS}}$, which is fully characterized by the correlation matrix $\varrho_{i j}=\rm{Tr}(\hat{\rho}_\rm{GS}\,\hat{c}^\dagger_{i} \hat{c}_{j})$. The spin index is absorbed into $i$ and $j$ for simplicity.
Considering the master equation \eqref{eq: master equation} and using Wick contractions, the EoM of the correlation matrix are given by
\begin{equation} \label{eq: eoms of GSA}
\begin{aligned}
        \dt \varrho_{ij}=&\langle i [\hat{H},\hat{c}_i^{\dagger} \hat{c}_j]- \frac{\gamma}{2}\{\hat{L}^\dag \hat{L}, \hat{c}_i^{\dagger} \hat{c}_j\} + \gamma \hat{L}^{\dagger}  \hat{c}_i^{\dagger} \hat{c}_j \hat{L} \rangle  \\
    =&-i \sum_k \langle \hat{c}_i^{\dagger} h_{j k} \hat{c}_k-\hat{c}_k^{\dagger} h_{k i} \hat{c}_j \rangle\\
    &-\frac{\gamma}{2}\langle \sum_k l_{j k} \hat{c}_i^{\dagger} \hat{c}_k+l_{k i} \hat{c}_k^{\dagger} \hat{c}_j\rangle,
\end{aligned}
\end{equation}
where $\langle \dots \rangle=\rm{Tr} (\hat{\rho}_\rm{GS} \dots)$ represents the expectation value and the Wick contraction theorem has been applied in the second equality. Since particle number is conserved, we focus exclusively on the Hartree-Fock contributions, i.e., we have discarded any anomalous correlation $\expval{\hat{c}_i \hat{c}_j}$ or $\expval*{\small \hat{c}_i^\dag \hat{c}_j^\dag}$.  The Hermitian matrices $h$ and $l$ are derived by expressing $\hat{H}=\hat{C}^\dag h \hat{C}$ and $(\hat{L}^\dag \hat{L})_\rm{HF} =\hat{C}^\dag l \hat{C}$, where $\hat{C}=(\hat{c}_{1},\hat{c}_{2},\hat{c}_{3},\ldots)^T$ denotes a column vector composed of the annihilation operators and $(\hat{L}^\dag \hat{L})_\rm{HF}$ represents the Hartree-Fock approximation of $\hat{L}^\dag \hat{L}$.

It can be shown that the EoM \eqref{eq: eoms of GSA} corresponds to an effective master equation involving one-body losses at the impurity site
\begin{equation}
    \dt\hat{\rho}_\rm{GS}(t)=-i\comm{\hat{H}}{\hat{\rho}_\rm{GS}}+\hat{\mathcal{L}}_\rm{GS}\hat{\rho}_\rm{GS},
\end{equation}
with a one-body dissipation term:
\begin{equation}    \hat{\mathcal{L}}_\rm{GS}\hat{\rho}_\rm{GS}=\sum_{\sigma\sigma'=\uparrow,\downarrow}\gamma_{\sigma\sigma'}(t)(\hat{c}_{0\sigma}\hat{\rho}_\rm{GS} \hat{c}_{0\sigma'}^\dag-\frac{1}{2}\{\hat{c}_{0\sigma'}^\dag \hat{c}_{0\sigma},\hat{\rho}_\rm{GS}\}),
\end{equation}
in which the one-body loss rates $\gamma_{\sigma\sigma'}(t) = \gamma\,\rm{Tr} [\hat{\rho}_\rm{GS}( t ) \hat{c}^\dag_{0\bar{\sigma}'}\hat{c}_{0\bar\sigma}]$ depend self-consistently on the correlation functions at the impurity sites. 

The effective master equation has the property that if the initial density matrix is a product state of the spin-up and spin-down sectors, such that $\hat{\rho}(t=0)=\hat{\rho}_\uparrow  \hat{\rho}_\downarrow$, this property will be conserved during the dynamics. Hence, the correlations $\varrho_{j\uparrow,j'\downarrow}=0$ will vanish throughout the evolution and we will have $\gamma_{\sigma\sigma^\prime}=\gamma \delta_{\sigma\sigma^\prime} n_{0\bar{\sigma}}(t)$. However, the Kondo effect at large dissipation implies that the true dynamics will generate a non-negligible entanglement between the two spin sectors. Hence, the GS approach cannot be valid beyond a moderate dissipation rate $\gamma\sim\Gamma$.

\bibliography{biblioDissKondo}

\end{document}